\documentclass[12pt,letter]{report}

\usepackage[pdftex]{graphicx} 
\usepackage{url} 
\usepackage[bookmarks, colorlinks=false, pdfborder={0 0 0}, pdftitle={<pdf title here>}, pdfauthor={<author's name here>}, pdfsubject={<subject here>}, pdfkeywords={<keywords here>}]{hyperref} 

\usepackage{multirow}
\usepackage{pdflscape}
\usepackage{makecell}
\usepackage{morefloats}
\usepackage{pdfpages}

\begin{document}
\renewcommand\bibname{References} 

\includepdf[pages={1}]{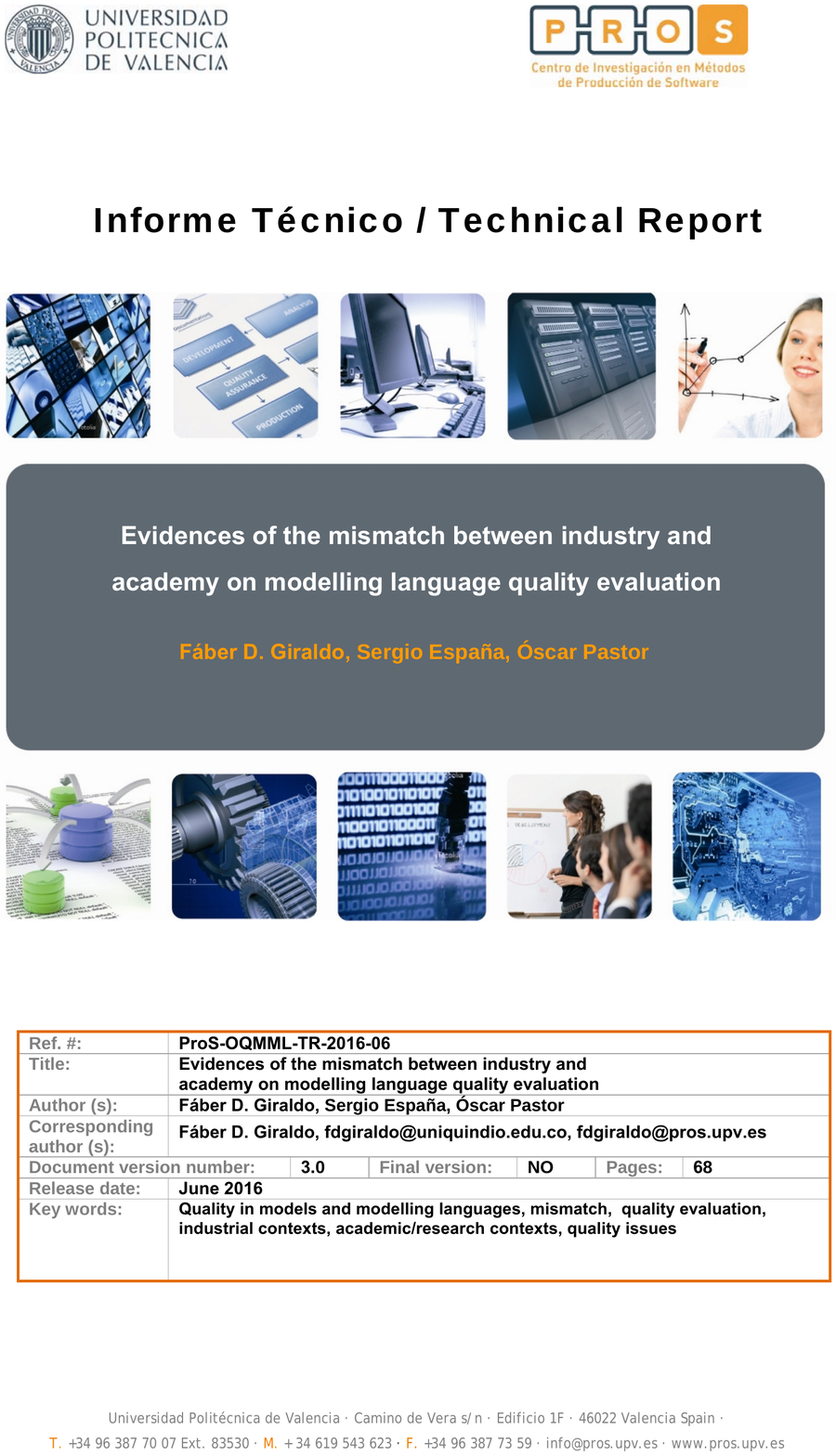}

\pagenumbering{roman} 
\tableofcontents
\listoftables

\newpage
\pagenumbering{arabic} 

\chapter{Introduction}

{\em Quality} is an implicit property of models and modelling languages by their condition of engineering artifacts. However, the quality property is affected by the diversity of conceptions around the {\em model-driven paradigm}. 

In this document is presented a report of quality issues on modelling languages and models. These issues result from an analysis about quality evidences obtained from industrial and academic/scientific contexts.

The found evidences are presented as follows:

\begin{itemize}
\item The citation of the work.
\item The year of publication.
\item The type of the work for each context (industrial and academic/research). The possible types for each context are: Journal papers, conference papers, workshop papers, technical reports, and web page (which include blogs, social network posts, forums, and similar). 
\item The detected issue.
\item The explicit sentences found in the work that support each detected quality issue.		
\end{itemize}

\chapter{Issues of MDE industrial practice relevant to modelling language quality evaluation}

This section presents several quality issues extracted from reports about MDE applicability in industrial practices. These quality issues impacts directly the perception of specific communities such as software developers and business/domain experts. Issues are reported with their associated works (paper, report, web page, and similar), the sources and their explicit mentions ({\em sentences}) around each quality issue.

\begin{landscape}


\clearpage
\newpage

\begin{table}[htbp]
  \centering
    \begin{tabular}{|p{2 cm}|p{1 cm}|p{2 cm}|p{3 cm}|p{10 cm}|}
      \hline
    \multicolumn{1}{|c|}{{\bf Source}} & \multicolumn{1}{|c|}{{\bf  Year}}  & \multicolumn{1}{|c|}{{\bf Type}} & \multicolumn{1}{|c|}{{\bf Issue}} & \multicolumn{1}{|c|}{{\bf Statements}} \\
    \hline

    \multirow{2}[2]{*}{\cite{Vara20122368}} & \multirow{2}[2]{*}{2012} & \multirow{3}[6]{*}{\parbox{2cm}{Journal paper}} & \multirow{5}[1]{*}{\parbox{3cm}{Implicit questions derived  from the MDE adoption itself }} & \multicolumn{1}{p{10 cm}|}{{\scriptsize {\em The problem of aligning high-level business models (corresponding to the business view) and information technologies (corresponding to the information system view) has become a crucial aspect in the field of software development.}}} \\

  &       &       &       &\\


  &       &       &       & \\ \hline
 
    \multirow{20}[6]{*}{{\cite{Cuadrado2014176}}} & \multirow{20}[6]{*}{2014} & \multirow{20}[6]{*}{\parbox{2cm}{Journal paper}} & \multirow{5}[1]{*}{\parbox{3cm}{Implicit questions derived from the MDE  adoption itself }} &  \multicolumn{1}{p{10 cm}|}{{\scriptsize {\em An important concern of any MDE project is its evolution as new requirements, new frameworks or new tools frequently appear.}}} \\

          &       &       &       & \multicolumn{1}{p{10 cm}|}{{\scriptsize {\em Normally both project managers and developers were initially attracted by the possibilities of MDE technology for automating software development, although they were skeptical about its applicability in real projects.}}}  \\

 &       &       &       &  \multicolumn{1}{p{10 cm}|}{{\scriptsize {\em Respondents consider that MDE requires them to carry out a significant extra training, and some interviewees have claimed that adopting MDE could have a significant risk as it could be difficult to have developers with an MDE education.}}}  \\

\cline{4-5}

 &       &       &  \multirow{5}[1]{*}{\parbox{3cm}{Tools as a way to increase complexity}}   & \multicolumn{1}{p{10 cm}|}{{\scriptsize {\em The evolution of the generated code has to be addressed to achieve a successful product and this is sometimes considered as one of the weak points in the adoption of MDE in industry.}}}   \\

  &       &       &       & \\ 

    \cline{4-5}

  & & & \multirow{5}[1]{*}{\parbox{3cm}{Implicit questions derived from the MDE  adoption itself }} &  \\
  &       &       &       & \\ 
  &       &       &       & \\ 

  &       &       &   \multirow{5}[1]{*}{\parbox{3cm}{\&}}    &   \multicolumn{1}{p{10 cm}|}{{\scriptsize {\em We have been reported that when facing a problem of translation between data formats  they considered using MDE techniques, but they were hold back because of the immaturity of the tools.}}}  \\ 

  &       &       &    \multirow{5}[1]{*}{\parbox{3cm}{Tools as a way to increase complexity}}    &\\

  &       &       &       & \\ 
  &       &       &       & \\   
&       &       &       & \\ 
    \hline

    \end{tabular}%
  \caption{Industrial issues evidenced (1/38).}
  \label{Table01}%
\end{table}%


\newpage

\begin{table}[htbp]
  \centering
    \begin{tabular}{|p{2 cm}|p{1 cm}|p{2 cm}|p{3 cm}|p{9 cm}|}
      \hline
    \multicolumn{1}{|c|}{{\bf Source}} & \multicolumn{1}{|c|}{{\bf  Year}}  & \multicolumn{1}{|c|}{{\bf Type}} & \multicolumn{1}{|c|}{{\bf Issue}} & \multicolumn{1}{|c|}{{\bf Statements}} \\
    \hline

        \multirow{20}[1]{*}{\cite{Brambilla201471}} & \multirow{20}[1]{*}{2014} & \multirow{20}[1]{*}{\parbox{2cm}{Journal paper}} & \multirow{5}[1]{*}{\parbox{2.5cm}{MDA is not enough}} & \multicolumn{1}{p{10 cm}|}{{\scriptsize {\em The manual implementation of the non-modeled parts of the application occurs independently of the model, which creates the well-known problems of 	model-to-code alignment; tracing changes from the model to the code and from the code to the models is overly complex and breaks most of the benefits of models during implementation and maintenance}}} \\

	&       &       &  &  \multicolumn{1}{p{10 cm}|}{{\scriptsize {\em MDE faces a barrier of mistrust in the quality of the generated code, which is inevitably perceived as less performing than the highly optimized code that developers can write using all the tricks of the trade.}}}  \\

	\cline{4-5}
          
	&       &       & \multirow{12}[1]{*}{\parbox{3cm}{Organizational support for the MDE adoption}} & \multicolumn{1}{p{10 cm}|}{{\scriptsize {\em Identifying the right role to discuss with: the main difficulty has been to identify the right role in the customer company to speak to. This may be a general problem for software vendors, but we think it is particularly critical for model-driven design tools: especially for large projects, the people in charge of the purchase decision power typically do not have the competencies and the capability for judging the quality and real impact of the tools, and thus rely on IT staff for an opinion. In turn, for IT people the attitude problem may play a role in the comments they provide.}}} \\
          &       &       &       & \multicolumn{1}{p{10 cm}|}{{\scriptsize {\em Motivating the company and the developers to address the learning curve: unfortunately, most of the IT staff do not have expertise and knowledge on MDE, therefore, the learning curve of the approach and of the specific modeling language is usually rather challenging and expensive.}}} \\

	\hline

        \multirow{8}[1]{*}{\cite{Lukman20131078}} & \multirow{8}[1]{*}{2013} & \multirow{8}[1]{*}{\parbox{2cm}{Journal paper}} & \multirow{5}[1]{*}{\parbox{2.5cm}{Organizational support for the MDE adoption}} & \multicolumn{1}{p{10 cm}|}{{\scriptsize {\em The problem is that these developers are reluctant to work in the early development activities and to use abstract high-level constructs, models and modeling languages, which are beneficial for taming the increased complexity of many process control systems.}}} \\
          &       &       &       & \multicolumn{1}{p{10 cm}|}{{\scriptsize {}}} \\ 
	\cline{4-5}

          &       &       &  \multirow{3}[1]{*}{\parbox{2cm}{MDA is not enough}} & \multicolumn{1}{p{10 cm}|}{{\scriptsize {\em The developers perceive GSMLs (e.g., general UML) as too vast, complex and inexpressive for the process control domain}}} \\
          &       &       &       & \multicolumn{1}{p{10 cm}|}{{\scriptsize {\em The majority of the available modeling languages are GPMLs.}}} \\
          &       &       &       & \multicolumn{1}{p{10 cm}|}{{\scriptsize {\em The problems of UML-PA are its vastness, the lack of automatic code generation, and lack of development guidelines/process.}}} \\

	\hline

        \hline
    \end{tabular}%
  \caption{Industrial issues evidenced (2/38).}
  \label{Table02}%
\end{table}%


\clearpage
\newpage

\begin{table}[htbp]
  \centering
    \begin{tabular}{|p{2 cm}|p{1 cm}|p{2 cm}|p{3 cm}|p{9 cm}|}
      \hline
    \multicolumn{1}{|c|}{{\bf Source}} & \multicolumn{1}{|c|}{{\bf  Year}}  & \multicolumn{1}{|c|}{{\bf Type}} & \multicolumn{1}{|c|}{{\bf Issue}} & \multicolumn{1}{|c|}{{\bf Statements}} \\
    \hline

       \multirow{7}[1]{*}{\cite{Hutchinson2014144}} & \multirow{7}[1]{*}{2014} & \multirow{7}[1]{*}{\parbox{2cm}{Journal paper}} & \multirow{5}[1]{*}{\parbox{2.5cm}{Organizational support for the MDE adoption}} & \multicolumn{1}{p{10 cm}|}{{\scriptsize {\em This case also highlights another difficulty for advocates of MDE — many users are employing techniques in entirely pragmatic 		ways with little or no interest in academic (or scientific) measurements of success and instead focus on their own organizational experience.}}} \\

          &       &       &  & \multicolumn{1}{p{10 cm}|}{{\scriptsize {\em Where MDE deployment seems to run into problems is where the decision to adopt an MDE approach is made without any real understanding of the necessary process change and instead takes on an autocratic top-down and wholesale character, 	implemented as an ‘‘all or nothing’’ approach.}}} \\

	 \cline{4-5}

          &       &       & \multirow{5}[1]{*}{\parbox{3cm}{Implicit questions derived from the MDE adoption itself
	}} & \multicolumn{1}{p{10 cm}|}{{\scriptsize {\em The answers about integrating generated code suggest much more ambivalence. About 36\% of respondents say it is not a significant problem, and a slightly higher 40\% say that it is a significant problem.}}} \\

          &       &       &  & \multicolumn{1}{p{10 cm}|}{{\scriptsize {\em MDE is not a ‘‘bolt on’’ process – i.e., it does not appear to offer benefits if it is simply added to existing processes. Instead, what is usually required is an overhaul of attitudes to certain aspects of how software should be developed.}}} \\

          &       &       &  & \multicolumn{1}{p{10 cm}|}{{\scriptsize {\em For example, code generation in MDE appears, at first glance, to have a positive effect on productivity. But the extra effort required to develop the models that make code generation possible, along with the possible need to make manual modifications, would appear to have a negative effect on productivity.}}} \\
    
        \hline
    \end{tabular}%
  \caption{Industrial issues evidenced (3/38).}
  \label{Table03}%
\end{table}%


\clearpage
\newpage

\begin{table}[htbp]
  \centering
    \begin{tabular}{|p{2 cm}|p{1 cm}|p{2 cm}|p{3 cm}|p{9 cm}|}
      \hline
    \multicolumn{1}{|c|}{{\bf Source}} & \multicolumn{1}{|c|}{{\bf  Year}}  & \multicolumn{1}{|c|}{{\bf Type}} & \multicolumn{1}{|c|}{{\bf Issue}} & \multicolumn{1}{|c|}{{\bf Statements}} \\
    \hline

       \multirow{7}[1]{*}{\cite{Hutchinson2014144}} & \multirow{7}[1]{*}{2014} & \multirow{7}[1]{*}{\parbox{2cm}{Journal paper}} & \multirow{5}[1]{*}{\parbox{2.5cm}{MDA is not enough}} & \multicolumn{1}{p{10 cm}|}{{\scriptsize {\em Since it was first introduced by the OMG in 1997, UML has quickly become the de-facto modeling language standard.Despite this, it has been widely criticized for its lack of a rigorous semantics3 [26], the process by which it has been ‘‘designed by committee’’, and its focus on graphical models (cf. [25]). Nevertheless, UML is widely used (as seen from the results in Section 2) and widely taught.}}} \\

	 \cline{4-5}

          &       &       & \multirow{5}[1]{*}{\parbox{3cm}{Tools as a way to increase complexity}} & \multicolumn{1}{p{10 cm}|}{{\scriptsize {\em How to interpret this diversity of tool usage is not obvious. It may suggest a rich array of tools to meet the different needs of users, or it may reflect an immaturity in a field where practitioners are still deciding on the best tools for the problem at hand- of users, or it may reflect an immaturity in a field where practitioners are still deciding on the best tools for the problem at hand.}}} \\

        \hline
    \end{tabular}%
  \caption{Industrial issues evidenced (4/38).}
  \label{Table04}%
\end{table}%


\newpage

\begin{table}[htbp]
  \centering
    \begin{tabular}{|p{2 cm}|p{1 cm}|p{2 cm}|p{3 cm}|p{9 cm}|}
      \hline
    \multicolumn{1}{|c|}{{\bf Source}} & \multicolumn{1}{|c|}{{\bf  Year}}  & \multicolumn{1}{|c|}{{\bf Type}} & \multicolumn{1}{|c|}{{\bf Issue}} & \multicolumn{1}{|c|}{{\bf Statements}} \\
    \hline

   	\multirow{7}[1]{*}{\cite{RePEc:rug:rugwps:07/462}}  & \multirow{7}[1]{*}{2014}  & \multirow{7}[1]{*}{\parbox{2cm}{Working paper}} & \multirow{7}[1]{*}{\parbox{2.5cm}{Tools as a way to increase complexity}}  & \multicolumn{1}{p{10 cm}|}{{\scriptsize {\em The use of a WE process with (semi-) automatic transformations prevents some development problems such as inconsistencies 			among models, lack of traceability, lack of technical soundness, etc. However, this (semi-)automatic nature of the WE process also may cause the propagation of quality flaws through levels of abstraction. Hence, quality problems that nowadays are just detected at deployment time may have been introduced at any previous stage of development.}}}  \\

        \hline

    	  \multirow{8}[1]{*}{\cite{Torchiano20132110}} & \multirow{8}[1]{*}{2013} & \multirow{8}[1]{*}{\parbox{2cm}{Journal paper}} & \multirow{5}[1]{*}{\parbox{2.5cm}{Tools as a way to increase complexity}} & \multicolumn{1}{p{10 cm}|}{{\scriptsize {\em They report that the biggest problem is the synchronization between models and code (models become out of date with code). We 	do not have evidence of this problem. Other problems are the quality of the generated code and issues with the modelling tools (e.g., too expensive, “heavyweight” and difficult to use).}}}  \\

	&       &       &       & \\

	\cline{4-5}
 
          &       &       & \multirow{5}[1]{*}{\parbox{2.5cm}{Organizational support for the MDE adoption}} & \multicolumn{1}{p{10 cm}|}{{\scriptsize {\em MDE is perceived as not simple: its complexity makes it viable for engineers but not for non technical people. This finding is confirmed by our results reported in Torchiano et al. (2011b) and Tomassetti et al. (2012). They show that only in a few 		cases business experts are involved during modelling tasks.}}} \\

	&       &       &       & \\

	\hline

    \end{tabular}%
  \caption{Industrial issues evidenced (5/38).}
  \label{Table05}%
\end{table}%


\newpage

\begin{table}[htbp]
  \centering
    \begin{tabular}{|p{2 cm}|p{1 cm}|p{2 cm}|p{3 cm}|p{9 cm}|}
      \hline
    \multicolumn{1}{|c|}{{\bf Source}} & \multicolumn{1}{|c|}{{\bf  Year}}  & \multicolumn{1}{|c|}{{\bf Type}} & \multicolumn{1}{|c|}{{\bf Issue}} & \multicolumn{1}{|c|}{{\bf Statements}} \\
    \hline

 \multirow{5}[1]{*}{\cite{Mohagheghi20131619}} & \multirow{5}[1]{*}{2013} & \multirow{5}[1]{*}{\parbox{2cm}{Journal paper}} & \multirow{5}[1]{*}{\parbox{2.5cm}{Organizational support for the MDE adoption}} & \multicolumn{1}{p{10 cm}|}{{\scriptsize {\em A language with too many concepts creates problems during language deployment and use.}}} \\

	&       &       &       & \\
	&       &       &       & \\
	&       &       &       & \\

    \hline

    \multirow{20}[1]{*}{\cite{BerLee:2008:Online}} & \multirow{20}[1]{*}{2009} & \multirow{20}[1]{*}{\parbox{2cm}{Web Site}} & \multirow{5}[1]{*}{\parbox{2.5cm}{MDA is not enough}} & \multicolumn{1}{p{10 cm}|}{{\scriptsize {\em This is still a tough problem to solve with a general solution.  For example as we move from a higher level model to a lower level model there is a fan out of elements – where a single element can spawn many elements at the lower level.  Then once created, the user can update, remove and add elements to the lower level model.}}} \\

          &       &       &  & \multicolumn{1}{p{10 cm}|}{{\scriptsize {\em Basically MDD was limited to a low level of abstraction and therefore could only have a limited impact. A lot of people simply used the MDD tools as a way of “visualizing” their code (think of a reverse transformation that gives you a graphical view of your code).}}} \\

	\cline{4-5}

          &       &       & \multirow{5}[1]{*}{\parbox{2.5cm}{Organizational support for the MDE adoption}} & \multicolumn{1}{p{10 cm}|}{{\scriptsize {\em Having the tools and infrastructure in place is not enough to deliver the software that addresses the challenging business issues we face. At the end, it is not the tools that solve the problem, but it is the people using the tools.}}} \\
          &       &       &       &  \multicolumn{1}{p{10 cm}|}{{\scriptsize {\em The right infrastructure and tools weren’t in place to reap the benefits of MDD.}}}  \\

	&       &       &       & \\

	\cline{4-5}

          &       &       & \multirow{7}[1]{*}{\parbox{2.5cm}{Implicit questions derived from the MDE adoption itself}} & \multicolumn{1}{p{10 cm}|}{{\scriptsize {\em It is very unlikely that a domain-independent tool will have built-in all the MDD artifacts that you require for your domain.}}} \\

	&       &       &       & \\
	&       &       &       & \\
	&       &       &       & \\
	&       &       &       & \\
	&       &       &       & \\

        \hline
    \end{tabular}%
  \caption{Industrial issues evidenced (6/38).}
  \label{Table06}%
\end{table}%


\newpage

\begin{table}[htbp]
  \centering
    \begin{tabular}{|p{2 cm}|p{1 cm}|p{2 cm}|p{3 cm}|p{9 cm}|}
      \hline
    \multicolumn{1}{|c|}{{\bf Source}} & \multicolumn{1}{|c|}{{\bf  Year}}  & \multicolumn{1}{|c|}{{\bf Type}} & \multicolumn{1}{|c|}{{\bf Issue}} & \multicolumn{1}{|c|}{{\bf Statements}} \\
    \hline

        \multirow{10}[1]{*}{\cite{Haan:2008:Online}} & \multirow{10}[1]{*}{2008} & \multirow{10}[1]{*}{\parbox{2cm}{Web Site}} & \multirow{7}[1]{*}{\parbox{2.5cm}{Implicit questions derived from the MDE adoption itself}} & \multicolumn{1}{p{10 cm}|}{{\scriptsize {\em One of the biggest pitfalls of MDE approaches is the cumbersome support for testing and debugging the software artifact on the model level.}}} \\
          &       &       &       & \multicolumn{1}{p{10 cm}|}{{\scriptsize {\em The two main quality criteria for models to be used in MDE are transformability and maintainability.}}} \\

	&       &       &       & \\
	&       &       &       & \\
	&       &       &       & \\

	\cline{4-5}

          &       &       & \multirow{3}[1]{*}{\parbox{2.5cm}{MDA is not enough}} & \multicolumn{1}{p{10 cm}|}{{\scriptsize {\em After the initial building process no support is delivered in managing the lifecycle of the software artifact. Changes have to be made in the generated source code or in parts of the model, this will lead to round-trip problems.}}} \\

          &       &       &  & \multicolumn{1}{p{10 cm}|}{{\scriptsize {\em MDA focusses on domain models. They define a CIM (computation independent model) which should be transformed into a PIM, which on its turn should be translated into a PSM. Problem is that a lot of definitions are going around, which is very confusing.}}} \\

	\cline{4-5}

          &       &       & \multirow{5}[1]{*}{\parbox{2.5cm}{Tools as a way to increase complexity}} & \multicolumn{1}{p{10 cm}|}{{\scriptsize {\em The last reason why MDE approaches (will) fail is the insufficient support of tooling. Tooling is essential to maximize the benefits of having models, and to minimize the effort required to maintain them  [1]}}} \\
          &       &       &       &  \multicolumn{1}{p{10 cm}|}{{\scriptsize {\em the tooling for an MDE approach should support teamwork for the development of big software projects. So, version control and support for distributed working are important.}}} \\

          &       &       &  & \multicolumn{1}{p{10 cm}|}{{\scriptsize {\em An MDE tool should provide a compiler-like behavior with error messages. These error message should identify problems in the models and not only within the generated source code.}}} \\

        \hline
    \end{tabular}%
  \caption{Industrial issues evidenced (7/38).}
  \label{Table07}%
\end{table}%


\newpage

\begin{table}[htbp]
  \centering
    \begin{tabular}{|p{2 cm}|p{1 cm}|p{2 cm}|p{3 cm}|p{9 cm}|}
      \hline
    \multicolumn{1}{|c|}{{\bf Source}} & \multicolumn{1}{|c|}{{\bf  Year}}  & \multicolumn{1}{|c|}{{\bf Type}} & \multicolumn{1}{|c|}{{\bf Issue}} & \multicolumn{1}{|c|}{{\bf Statements}} \\
    \hline

   \multirow{20}[1]{*}{\cite{6507223}} & \multirow{20}[1]{*}{2014} & \multirow{20}[1]{*}{\parbox{2cm}{Journal Paper}} & \multirow{5}[1]{*}{\parbox{2.5cm}{Organizational support for the MDE adoption}} & \multicolumn{1}{p{10 cm}|}{{\scriptsize {\em Even in companies that recognize the benefits of MDE, adoption can take a long time, even when compared to the adoption of other approaches such as agile. Our data illustrates that one of the main factors for this inertia is that MDE is usually marketed as a technology that can do the same things faster and cheaper.}}}  \\

          &       &       &  & \multicolumn{1}{p{10 cm}|}{{\scriptsize {\em In addition, it appears that MDE developers need both compiler development skills and abstraction skills. Unfortunately, these skillsets are usually taught in distinct parts of a computer science curriculum with little connection between them.}}} \\
          &       &       &       &  \multicolumn{1}{p{10 cm}|}{{\scriptsize {\em A recent study surveyed 50 software designers and found that these designers either didn't use UML at all or used it only selectively and informally.}}} \\
          &       &       &       & \multicolumn{1}{p{10 cm}|}{{\scriptsize {\em UML 2.0, for example, a major revision of the UML standard, didn’t reflect the literature on empirical studies of software modeling or software design studies. Consequently, current approaches force developers and organizations to operate in a way that fi ts the approach instead of making the approach fit the people.}}} \\

	\cline{4-5}

          &       &       & \multirow{7}[1]{*}{\parbox{2.5cm}{Implicit questions derived from the MDE adoption itself}} & \multicolumn{1}{p{10 cm}|}{{\scriptsize {\em Companies would therefore be wise to consider the more holistic benefits that MDE can bring rather than focusing only on code generation.}}} \\

	&       &       &       & \\
	&       &       &       & \\
	&       &       &       & \\
	&       &       &       & \\
	&       &       &       & \\

        \hline
    \end{tabular}%
  \caption{Industrial issues evidenced (8/38).}
  \label{Table08}%
\end{table}%


\newpage

\begin{table}[htbp]
  \centering
    \begin{tabular}{|p{2 cm}|p{1 cm}|p{2 cm}|p{3 cm}|p{9 cm}|}
      \hline
    \multicolumn{1}{|c|}{{\bf Source}} & \multicolumn{1}{|c|}{{\bf  Year}}  & \multicolumn{1}{|c|}{{\bf Type}} & \multicolumn{1}{|c|}{{\bf Issue}} & \multicolumn{1}{|c|}{{\bf Statements}} \\
    \hline

 \multirow{25}[1]{*}{\cite{JeJac:2011:Online}} & \multirow{25}[1]{*}{2011} & \multirow{25}[1]{*}{\parbox{2cm}{Web Site}} & \multirow{7}[1]{*}{\parbox{2.5cm}{Implicit questions derived from the MDE adoption itself}} & \multicolumn{1}{p{10 cm}|}{{\scriptsize {\em There has not been any killer application that produces measurable and reproducible evidence that MDE provides at least an order of magnitude improvement over previous solutions}}} \\
          &       &       &       & \multicolumn{1}{p{10 cm}|}{{\scriptsize {\em There are too many definitions and camps (MDD, MDSD, MDA, MDE, MDSE, MBD,...)}}} \\
          &       &       &       & \multicolumn{1}{p{10 cm}|}{{\scriptsize {\em There is still a lot of confusion about what MDE is, for instance, between Simulation and MDE, two very different branches of software engineering}}} \\
          &       &       &       & \multicolumn{1}{p{10 cm}|}{{\scriptsize {\em UML, which is often use as a foundation to modeling, is a loosely defined langugage, built by industrial consensus, with poor modularity principles, too big, too complex, changes too often,…}}} \\

 &  &  & & \multicolumn{1}{p{10 cm}|}{{\scriptsize {\em MDE lacks modularity, which was not solved by the introduction of packages and profiles in UML.}}} \\

  &  &   &   & \multicolumn{1}{p{10 cm}|}{{\scriptsize {\em MDE lacks portability: there are no serious results on portability of models in time and space (a model produced 10 years ago, may not be exploitable today).}}} \\

  &  &   &   & \multicolumn{1}{p{10 cm}|}{{\scriptsize {\em MDE focused too much on the model of code, and not enough on the model of data
MDE focused too much on solution models and not enough on problem models MDE focused too much on Information System models and not enough on Business Models MDE focused too much on modeling in the small and not enough in the large.}}} \\

          &       &       &  & \multicolumn{1}{p{10 cm}|}{{\scriptsize {\em  MDE is often perceived as adding complexity: Metamodels are often too large, there are too many, and their relationships and alignment are poorly understood}}}\\

        \hline
    \end{tabular}%
  \caption{Industrial issues evidenced (9/38).}
  \label{Table09}%
\end{table}%


\newpage

\begin{table}[htbp]
  \centering
    \begin{tabular}{|p{2 cm}|p{1 cm}|p{2 cm}|p{3 cm}|p{9 cm}|}
      \hline
    \multicolumn{1}{|c|}{{\bf Source}} & \multicolumn{1}{|c|}{{\bf  Year}}  & \multicolumn{1}{|c|}{{\bf Type}} & \multicolumn{1}{|c|}{{\bf Issue}} & \multicolumn{1}{|c|}{{\bf Statements}} \\
    \hline

 \multirow{12}[1]{*}{\cite{JeJac:2011:Online}} & \multirow{12}[1]{*}{2011} & \multirow{12}[1]{*}{\parbox{2cm}{Web Site}} & \multirow{7}[1]{*}{\parbox{2.5cm}{Organizational support for the MDE adoption}} & \multicolumn{1}{p{10 cm}|}{{\scriptsize {\em There is a confusion between programming and modeling that emerged in the 90s when visual programming languages appeared.}}} \\

	&       &       &       & \\
	&       &       &       & \\
	&       &       &       & \\
	&       &       &       & \\
	\cline{4-5}

          &       &       & \multirow{3}[1]{*}{\parbox{2.5cm}{MDA is not enough}} & \multicolumn{1}{p{10 cm}|}{{\scriptsize {\em Execution vs Precision: precision is not always obtained through execution, and MDE does not always aim at creating executable models}}} \\

      &       &       & & \multicolumn{1}{p{10 cm}|}{{\scriptsize {\em The notion of a platform model was never taken seriously, and the CIM/PIM/PSM was a false good idea.}}} \\

          &       &       &  & \multicolumn{1}{p{10 cm}|}{{\scriptsize {\em XMI was a failure, it will eventually disappear, creating a maintenance problem for UML.}}} \\

   &       &       &  & \multicolumn{1}{p{10 cm}|}{{\scriptsize {\em UML, the main problem Confusing modeling and programming languages
No definition / evaluation of executable modeling Too many camps.}}} \\

        \hline
    \end{tabular}%
  \caption{Industrial issues evidenced (10/38).}
  \label{Table10}%
\end{table}%


\newpage

\begin{table}[htbp]
  \centering
    \begin{tabular}{|p{2 cm}|p{1 cm}|p{2 cm}|p{3 cm}|p{9 cm}|}
      \hline
    \multicolumn{1}{|c|}{{\bf Source}} & \multicolumn{1}{|c|}{{\bf  Year}}  & \multicolumn{1}{|c|}{{\bf Type}} & \multicolumn{1}{|c|}{{\bf Issue}} & \multicolumn{1}{|c|}{{\bf Statements}} \\
    \hline

     \multirow{10}[1]{*}{\cite{Ben:2015:Online}} & \multirow{10}[1]{*}{2015} & \multirow{10}[1]{*}{\parbox{2cm}{Web Site}} & \multirow{9}[1]{*}{\parbox{2.5cm}{Implicit questions derived from the MDE adoption itself}} & \multicolumn{1}{p{10 cm}|}{{\scriptsize {\em MDSE does not yet offer a holistic, homogenous solution for software engineering and probably won’t for another 10 years or more.}}} \\

&  &  & &  \multicolumn{1}{p{10 cm}|}{{\scriptsize {\em The simple fact that there is no consistent name for the subject – it’s referred to as {\em model based software development}, {\em model driven software engineering (MDSE)}, {\em model driven development}, etc –indicates that there is a lack of clarity about what exactly is meant by the term.}}}  \\

          &       &       &       &\multicolumn{1}{p{10 cm}|}{{\scriptsize {\em So, specifically, I think that MDSE offers solutions of proven value for the majority of aspects of the design of algorithms (e.g. Matlab) and discrete control (e.g., ASD:Suite/Dezyne)}}}\\

    \cline{4-5}

          &       &       & \multirow{7}[1]{*}{\parbox{2.5cm}{Tools as a way to increase complexity}} & \multicolumn{1}{p{10 cm}|}{{\scriptsize {\em At the moment there are few companies making use of data modelling techniques, in principle because the tool offering in this area is thin and offers little value.}}} \\
	&       &       &       & \\
	&       &       &       & \\
	&       &       &       & \\
	&       &       &       & \\  

        \hline
    \end{tabular}%
  \caption{Industrial issues evidenced (11/38).}
  \label{Table11}%
\end{table}%


\newpage

\begin{table}[htbp]
  \centering
    \begin{tabular}{|p{2 cm}|p{1 cm}|p{2 cm}|p{3 cm}|p{9 cm}|}
      \hline
    \multicolumn{1}{|c|}{{\bf Source}} & \multicolumn{1}{|c|}{{\bf  Year}}  & \multicolumn{1}{|c|}{{\bf Type}} & \multicolumn{1}{|c|}{{\bf Issue}} & \multicolumn{1}{|c|}{{\bf Statements}} \\
    \hline

 \multirow{20}[1]{*}{\cite{Lars:2008:Online}} & \multirow{20}[1]{*}{2008} & \multirow{20}[1]{*}{\parbox{2cm}{Web Site}} & \multirow{5}[1]{*}{\parbox{2.5cm}{Tools as a way to increase complexity}} & \multicolumn{1}{p{10 cm}|}{{\scriptsize {\em Modelling tools might suffer from problems using versioning tools. Source code is usually simpler to merge than a model diagram. This forces a team to move from the copy-edit-merge to a lock-edit-merge workflow.}}} \\

          &       &       & &  \multicolumn{1}{p{10 cm}|}{{\scriptsize {\em Sometimes it means turning UML or another type of diagrams in to executable code. I've never seen this work out well with the tools available nowadays.}}} \\
          &       &       &       & \multicolumn{1}{p{10 cm}|}{{\scriptsize {\em It usually causes projects to get results really fast and then cause a maintanance nightmare because the tools available don't really support big teams working on visual diagrams and because people start working in the diagrams as well as the generated code.}}} \\

    \cline{4-5}

          &       &       & \multirow{5}[1]{*}{\parbox{2.5cm}{MDA is not enough}} & \multicolumn{1}{p{10 cm}|}{{\scriptsize {\em If you have a model, you want every line of code to come from that model. And it may be difficult to include external libraries to a project. So either you live with the fact, that your system is based on external components or you reinvent the wheel to get it into your model.}}} \\
          &       &       &       & \multicolumn{1}{p{10 cm}|}{{\scriptsize {\em MDA is a bit of an overloaded concept.}}} \\

          &       &       &   & \multicolumn{1}{p{10 cm}|}{{\scriptsize {\em The problem with actual application aiming at the three tiered MDA approach is that those are terribly difficult to set up and adapt to specific requirements. Just think of ABAP and SAP.}}} \\
          &       &       &       & \multicolumn{1}{p{10 cm}|}{{\scriptsize {\em MDA usually make difficult to integrate the business rules inside the server side layer.}}} \\
  
        \hline
    \end{tabular}%
  \caption{Industrial issues evidenced (12/38).}
  \label{Table12}%
\end{table}%


\newpage

\begin{table}[htbp]
  \centering
    \begin{tabular}{|p{2 cm}|p{1 cm}|p{2 cm}|p{3 cm}|p{9 cm}|}
      \hline
    \multicolumn{1}{|c|}{{\bf Source}} & \multicolumn{1}{|c|}{{\bf  Year}}  & \multicolumn{1}{|c|}{{\bf Type}} & \multicolumn{1}{|c|}{{\bf Issue}} & \multicolumn{1}{|c|}{{\bf Statements}} \\
    \hline

    \multirow{7}[1]{*}{\cite{Tone:2010:Online}} & \multirow{7}[1]{*}{2010} & \multirow{7}[1]{*}{\parbox{2cm}{Web Site}} & \multirow{7}[1]{*}{\parbox{2.5cm}{Implicit questions derived from the MDE adoption itself}} & \multicolumn{1}{p{10 cm}|}{{\scriptsize {\em We did that once with a 3 mloc logistics planner system, and it worked well. However, we realized early on that 			UML would not be sufficient. It was simply too obtuse to capture the level of detail needed for the specification.}}} \\
          &       &       &       & \multicolumn{1}{p{10 cm}|}{{\scriptsize {\em There are many things which might go wrong, like e.g. editing generated code directly, being able to generate only once, becaause manually edited code would be erased after generation}}} \\

	&       &       &       & \\  

        \hline

    \multirow{7}[1]{*}{\cite{Finnie2015}} & \multirow{7}[1]{*}{2015} & \multirow{7}[1]{*}{\parbox{2cm}{Web Site}} & \multirow{7}[1]{*}{\parbox{2.5cm}{Implicit questions derived from the MDE adoption itself}} & \multicolumn{1}{p{10 cm}|}{{\scriptsize {\em So what are we missing in the MDx community? There may be several factors. But two stand out above all the others: We forgot why we model; We forgot who models. These are kind of important. In fact, stark-raving fundamentally critical. No wonder we’re in a pickle.}}} \\

	&       &       &       & \\  

	&       &       &       & \\  

	&       &       &       & \\

\hline
    \end{tabular}%
  \caption{Industrial issues evidenced (13/38).}
  \label{Table13}%
\end{table}%


\newpage

\begin{table}[htbp]
  \centering
    \begin{tabular}{|p{2 cm}|p{1 cm}|p{2 cm}|p{3 cm}|p{9 cm}|}
      \hline
    \multicolumn{1}{|c|}{{\bf Source}} & \multicolumn{1}{|c|}{{\bf  Year}}  & \multicolumn{1}{|c|}{{\bf Type}} & \multicolumn{1}{|c|}{{\bf Issue}} & \multicolumn{1}{|c|}{{\bf Statements}} \\
    \hline

    \multirow{25}[1]{*}{\cite{Martin:2008:Online}} & \multirow{25}[1]{*}{2009} & \multirow{25}[1]{*}{\parbox{2cm}{Web Site}} & \multirow{5}[1]{*}{\parbox{2.5cm}{Tools as a way to increase complexity}} & \multicolumn{1}{p{10 cm}|}{{\scriptsize {\em I think, it will take time, till the tools get more refined, more people gain experience with MDD. At the moment if you want to get something out of MDD you have to invest quite a lot, so its use remains limited. }}}\\

        &       &       &  & \multicolumn{1}{p{10 cm}|}{{\scriptsize {\em Of course popularity isn't everything, and things to have a tendency to come back, but for the time being I think MDA+tools is viewed by many as "wizard based code generation" tools (regardless of what it really is) so I think it will be some time or perhaps never that it really takes off.}}} \\

\cline{4-5}

          &       &       & \multirow{3}[1]{*}{\parbox{2.5cm}{MDA is not enough}} & \multicolumn{1}{p{10 cm}|}{{\scriptsize {\em I think MDSD is still too much tied to code generation.}}} \\

	&       &       &       & \\
	&       &       &       & \\  

\cline{4-5}

          &       &       & \multirow{5}[1]{*}{\parbox{2.5cm}{Organizational support for the MDE adoption}} & \multicolumn{1}{p{10 cm}|}{{\scriptsize {\em One of the problems of MDD is that, since it works on an higher abstraction level, it requires developers that can go up on the abstraction level too. That greatly reduces the universe of developers who can understand and use such methodologies.}}} \\

	&       &       &       & \\ 

\cline{4-5}

          &       &       & \multirow{8}[1]{*}{\parbox{2.5cm}{Implicit questions derived from the MDE adoption itself}} & \multicolumn{1}{p{10 cm}|}{{\scriptsize {\em Model Driven Development will be the future if and only if the models that it uses can be as flexible as writing the code that it's supposed to be generating. I think the reason why it's not doing so well right now is that you it's difficult to do the same "round-tripping" that text-based programming languages have been doing for decades.}}} \\
          &       &       &       &  \multicolumn{1}{p{10 cm}|}{{\scriptsize {\em I can't quite put my finger on it, but there's still something missing in MDD that would make it as useful as people would claim it to be.}}} \\
          &       &       &       & \multicolumn{1}{p{10 cm}|}{{\scriptsize {\em IMHO, the only way MDD can ever be useful if it's built from the ground up to be as expressive and as flexible as its text-based counterpart.}}} \\

        \hline
    \end{tabular}%
  \caption{Industrial issues evidenced (14/38).}
  \label{Table14}%
\end{table}%


\newpage

\begin{table}[htbp]
  \centering
    \begin{tabular}{|p{2 cm}|p{1 cm}|p{2 cm}|p{3 cm}|p{9 cm}|}
      \hline
    \multicolumn{1}{|c|}{{\bf Source}} & \multicolumn{1}{|c|}{{\bf  Year}}  & \multicolumn{1}{|c|}{{\bf Type}} & \multicolumn{1}{|c|}{{\bf Issue}} & \multicolumn{1}{|c|}{{\bf Statements}} \\
    \hline

 \multirow{27}[1]{*}{\cite{Kulkarni20106395}} & \multirow{27}[1]{*}{2010} & \multirow{27}[1]{*}{\parbox{2cm}{Conference Paper}} & \multirow{5}[1]{*}{\parbox{2.5cm}{Organizational support for the MDE adoption}} & \multicolumn{1}{p{10 cm}|}{{\scriptsize {\em Large business applications are typically implemented using distributed architecture. This requires expertise in a variety of technologies such as graphical user interface frameworks, middleware platforms, databases and a variety of programming languages.}}} \\
          &       &       &       & \multicolumn{1}{p{10 cm}|}{{\scriptsize {\em Architecture expertise is in short supply, and conceiving the right application architecture and ensuring that it is implemented correctly and efficiently by a large team of developers is a difficult problem.}}}  \\

          &       &       &  & \multicolumn{1}{p{10 cm}|}{{\scriptsize {\em Large business applications require large development teams. Typically, development effort is partitioned along functional modules that have high internal cohesion and low external coupling. In absence of explicitly stated dependencies, such a partitioning may introduce spurious dependencies that can lead to integration problems.}}} \\

\cline{4-5}

          &       &       & \multirow{4}[1]{*}{\parbox{2.5cm}{MDA is not enough}} & \multicolumn{1}{p{10 cm}|}{{\scriptsize {\em As these models are the principal artefacts in SDLC, they must be first class entities, intuitive and closer to the problem domain. We found that by force-fitting all models into UML or UML stereotypes, these important properties are lost.}}}  \\

\cline{4-5}

          &       &       & \multirow{8}[1]{*}{\parbox{2.5cm}{Tools as a way to increase complexity}} & \multicolumn{1}{p{10 cm}|}{{\scriptsize {\em Ensuring complete and consistent implementation of such a change became difficult as it required thorough knowledge of all the concerned code generators on the part of a tool implementer. This problem became more acute as the number of variants of the code generators grew.}}}   \\

          &       &       &  & \multicolumn{1}{p{10 cm}|}{{\scriptsize {\em Most MDE tools available wouldn't have scaled up to our needs as they are based on single-user file-based storage for models. These tools do provide basic mechanisms to organize models into containers and code into directory structures.}}}  \\

          &       &       &	 & \multicolumn{1}{p{10 cm}|}{{\scriptsize {\em Most MDE tools do not come with a change model inbuilt. As a result, the onus of determining what has changed in a model, what impact the change has on other model and code artefacts, and how to propagate the change correctly to all the impacted artefacts lies entirely with the developer. This is an error-prone and effort intensive activity.}}} \\
    \hline

        \hline
    \end{tabular}%
  \caption{Industrial issues evidenced (15/38).}
  \label{Table15}%
\end{table}%


\newpage

\begin{table}[htbp]
  \centering
    \begin{tabular}{|p{2 cm}|p{1 cm}|p{2 cm}|p{3 cm}|p{9 cm}|}
      \hline
    \multicolumn{1}{|c|}{{\bf Source}} & \multicolumn{1}{|c|}{{\bf  Year}}  & \multicolumn{1}{|c|}{{\bf Type}} & \multicolumn{1}{|c|}{{\bf Issue}} & \multicolumn{1}{|c|}{{\bf Statements}} \\
    \hline

 \multirow{15}[1]{*}{\cite{Antonio2014}} & \multirow{15}[1]{*}{2014} & \multirow{15}[1]{*}{\parbox{2cm}{Journal Paper}} & \multirow{9}[1]{*}{\parbox{2.5cm}{Implicit questions derived from the MDE adoption itself}} & \multicolumn{1}{p{10 cm}|}{{\scriptsize {\em Regarding the use of design models in industry, all systematic studies seem to agree that the knowledge about modeling and its use are not widespread yet. In the main, developers do not make the best use of models and tend to perceive little or no value added in modeling.}}} \\

          &       &       &  & \multicolumn{1}{p{10 cm}|}{{\scriptsize {\em Many software engineers currently use diagrams and simulations in their work but do not consider they are modeling.}}} \\

          &       &       &  & \multicolumn{1}{p{10 cm}|}{{\scriptsize {\em developers are not really interested in another language which is restricted to constraint expressions. The benefits of models are not seen in comparison to programming languages
by many current software developers.}}} \\

          &       &       &  & \multicolumn{1}{p{10 cm}|}{{\scriptsize {\em Conservative mindset of many software practitioners; resistance to technological change, even if the new technology can lead to better results.}}} \\

    \cline{4-5}

 &  & &\multirow{5}[1]{*}{\parbox{2.5cm}{Tools as a way to increase complexity}} & \multicolumn{1}{p{10 cm}|}{{\scriptsize {\em Tool support is still insufficient, in particular for model validation, simulation and interchange; for specifying constraints and correspondences between models;}}} \\

	&       &       &       & \\  
	&       &       &       & \\

\cline{4-5}

          &       &       & \multirow{5}[1]{*}{\parbox{2.5cm}{MDA is not enough}} & \multicolumn{1}{p{10 cm}|}{{\scriptsize {\em Many software practitioners are still completely unaware of modeling notations and of MDE.}}} \\

          &       &       &  & \multicolumn{1}{p{10 cm}|}{{\scriptsize {\em One of them is that MDE involves dependent activities that have both positive and negative effects. For example, automatic code generation can have a positive effect on productivity. But the extra effort required to develop the models, along with the possible need to make manual modifications, may have a negative effect. The balance between these two effects is related to context, and needs to be carefully considered}}} \\

    \hline

    \end{tabular}%
  \caption{Industrial issues evidenced (16/38).}
  \label{Table16}%
\end{table}%


\newpage

\begin{table}[htbp]
  \centering
    \begin{tabular}{|p{2 cm}|p{1 cm}|p{2 cm}|p{3 cm}|p{9 cm}|}
      \hline
    \multicolumn{1}{|c|}{{\bf Source}} & \multicolumn{1}{|c|}{{\bf  Year}}  & \multicolumn{1}{|c|}{{\bf Type}} & \multicolumn{1}{|c|}{{\bf Issue}} & \multicolumn{1}{|c|}{{\bf Statements}} \\
    \hline

 \multirow{25}[1]{*}{\cite{Antonio2014}} & \multirow{25}[1]{*}{2014} &  \multirow{25}[1]{*}{\parbox{2cm}{Journal Paper}} & \multirow{15}[1]{*}{\parbox{2.5cm}{Organizational support for the MDE adoption}} & \multicolumn{1}{p{10 cm}|}{{\scriptsize {\em Lack of education, team experience and skills sets in most developers and software practitioners. MDE is not only a change in technology, but a complete paradigm shift.}}}  \\
          &       &       &       & \multicolumn{1}{p{10 cm}|}{{\scriptsize {\em Too much emphasis on technology and not enough on technology users and their needs.}}} \\
          &       &       &       & \multicolumn{1}{p{10 cm}|}{{\scriptsize {\em Inadequate or flawed information about MDE concepts, goals, tools and real achievements – for many companies it is not clear whether MDE is just an academic theory, the tool vendor’s sales pitch or if there are indeed many organizations successfully using MDE to realize measurable benefits on real software engineering projects. − Lack of “systems” perspective and lack of abstraction skills.}}} \\

          &       &       &  & \multicolumn{1}{p{10 cm}|}{{\scriptsize {\em Design models are not used very extensively, and where they are used, the use is informal and mostly without tool support; the notation is often not UML but many others.}}} \\

          &       &       &  & \multicolumn{1}{p{10 cm}|}{{\scriptsize {\em Models are seldom updated after initial creation, and are usually drawn on a white board or on paper.}}} \\

          &       &       &  & \multicolumn{1}{p{10 cm}|}{{\scriptsize {\em Current business climate heavily focused on short-term gain discourages investment in new methods and tools.}}} \\
          &       &       &       & \multicolumn{1}{p{10 cm}|}{{\scriptsize {\em Development teams’ re-education and training can indeed be expensive (since it may imply
changing their mindsets, not only their methods and tools).}}} \\

          &       &       &  & \multicolumn{1}{p{10 cm}|}{{\scriptsize {\em Modeling also requires a mindset change. The use of (the right) abstraction techniques is more difficult and less common than expected [SZ13], and modeling requires different skills other than programming.}}} \\
          &       &       &       & \multicolumn{1}{p{10 cm}|}{{\scriptsize {\em One of the major reasons professional practitioners give for declining to use UML is that, after due consideration, they have concluded that it doesn’t add sufficient value to warrant the cost of transition. Many practitioners already have a repertoire of tools and representations that has been thoughtfully developed and evolved over time to fit their effective practices.}}} \\

    \hline
    \end{tabular}%
  \caption{Industrial issues evidenced (17/38).}
  \label{Table17}%
\end{table}%


\newpage

\begin{table}[htbp]
  \centering
    \begin{tabular}{|p{2 cm}|p{1 cm}|p{2 cm}|p{3.5 cm}|p{9 cm}|}
      \hline
    \multicolumn{1}{|c|}{{\bf Source}} & \multicolumn{1}{|c|}{{\bf  Year}}  & \multicolumn{1}{|c|}{{\bf Type}} & \multicolumn{1}{|c|}{{\bf Issue}} & \multicolumn{1}{|c|}{{\bf Statements}} \\
    \hline

 \multirow{15}[1]{*}{\cite{Hebig:2014:NSI:2600821.2600846}} & \multirow{15}[1]{*}{2014} & \multirow{15}[1]{*}{\parbox{2cm}{Conference Paper}} & \multirow{15}[1]{*}{\parbox{3.5cm}{Implicit questions derived from the MDE adoption itself}} &  \multicolumn{1}{p{10 cm}|}{{\scriptsize {\em We come up with the observation that - although it is in some cases possible to reuse standard processes - the combination with MDE can also result in heavyweight changes to a process.}}}  \\
          &       &       &       &   \multicolumn{1}{p{10 cm}|}{{\scriptsize {\em there is still a lack of systematic knowledge on the impact of MDE on software processes and on necessary adaptation steps for combining processes with MDE.}}} \\
          &       &       &       &  \multicolumn{1}{p{10 cm}|}{{\scriptsize {\em the results show that MDE settings can have a higher impact on process tailoring than expected. While single factors such as used languages and tool infrastructure are already considered ([8]), the composition with MDE settings is not taken into account by most process tailoring approaches.}}}  \\
          &       &       &       &  \multicolumn{1}{p{10 cm}|}{{\scriptsize {\em the process adaptations that we could observe (e.g. changes in the structure of phases) go much further than the tailoring actions that have been collected by Kalus et al. [8] (e.g. increasing the number of (micro-)iterations).}}} \\
          &       &       &       &  \multicolumn{1}{p{10 cm}|}{{\scriptsize {\em It seems that the role of MDE for process tailoring is underestimated, so far. There is a lack of guidance for process managers who need to cope with an introduction of MDE.}}}
 \\
    \hline

 \multirow{2}[1]{*}{\cite{DenHaan:2010:Online}} & \multirow{2}[1]{*}{2010} & \multirow{2}[1]{*}{\parbox{2cm}{Web site}} & \multirow{2}[1]{*}{\parbox{3.5cm}{Tools as a way to increase complexity}} &  \multicolumn{1}{p{10 cm}|}{{\scriptsize {\em The second release cycle joins the party when the MDE tool doesn’t support the changes you need to make to your application.}}}  \\
    \hline

    \end{tabular}%
  \caption{Industrial issues evidenced (18/38).}
  \label{Table18}%
\end{table}%


\newpage

\begin{table}[htbp]
  \centering
    \begin{tabular}{|p{2 cm}|p{1 cm}|p{2 cm}|p{3 cm}|p{9 cm}|}
      \hline
    \multicolumn{1}{|c|}{{\bf Source}} & \multicolumn{1}{|c|}{{\bf  Year}}  & \multicolumn{1}{|c|}{{\bf Type}} & \multicolumn{1}{|c|}{{\bf Issue}} & \multicolumn{1}{|c|}{{\bf Statements}} \\
    \hline

 \multirow{15}[1]{*}{\cite{Brown:2009:Online}} & \multirow{15}[1]{*}{2009} & \multirow{15}[1]{*}{\parbox{2cm}{Web site}} & \multirow{3}[1]{*}{\parbox{2.5cm}{MDA is not enough}} &  \multicolumn{1}{p{10 cm}|}{{\scriptsize {\em As I read over the literature about MDD and MDA over the years, I am amused how it turned out to be another factor favoring complexity instead of diminishing it.}}}  \\
    \cline{5-5}

 & & & & {{\scriptsize {\em OMG’s MDA specifications are not usable by every software company: Small software companies can’t afford to implement specifications such as MOF, UML, XMI and CWM. PIM and PSM should be only used as an abstraction layer not a specification set.}}}  \\
    \cline{4-5}

 & & & \multirow{7}[1]{*}{\parbox{2.5cm}{Implicit questions derived from the MDE adoption itself}} & {{\scriptsize {\em As software builders, we must be much more result-oriented and focus on utilizing concepts (Principles) rather than getting lost in specifications (Specification evaluation is another topic). }}}  \\

          &       &       &       &\\
          &       &       &       & \\
          &       &       &       &\\
          &       &       &       &\\
   \hline

    \end{tabular}%
  \caption{Industrial issues evidenced (19/38).}
  \label{Table19}%
\end{table}%


\newpage

\begin{table}[htbp]
  \centering
    \begin{tabular}{|p{2 cm}|p{1 cm}|p{2 cm}|p{3 cm}|p{9 cm}|}
      \hline
    \multicolumn{1}{|c|}{{\bf Source}} & \multicolumn{1}{|c|}{{\bf  Year}}  & \multicolumn{1}{|c|}{{\bf Type}} & \multicolumn{1}{|c|}{{\bf Issue}} & \multicolumn{1}{|c|}{{\bf Statements}} \\
    \hline

 \multirow{7}[1]{*}{\cite{DenHaan:2009:Online}} & \multirow{7}[1]{*}{2009} & \multirow{7}[1]{*}{\parbox{2cm}{Web site}} & \multirow{7}[1]{*}{\parbox{2.5cm}{Implicit questions derived from the MDE adoption itself}} &  \multicolumn{1}{p{10 cm}|}{{\scriptsize {\em If you’re used to programming everything by hand, MDD can be quite rigid. The goal of MDD is to ‘program’ on a higher level of abstraction.}}}  \\

 & & & & {{\scriptsize {\em The problem with using models to directly drive the engineering of software is that they are far from flexible. }}}  \\
 & & & & {{\scriptsize {\em  Limited by the kind of Model Driven Engineering tool you use}}}  \\
 & & & & {{\scriptsize {\em You’re only flexible in the parts of the solution covered by the used Domain-Specific Languages.}}}  \\
 & & & & {{\scriptsize {\em If the factory isn’t finished and tested in practice before you are going to start a big project you will have a huge risk}}}  \\
 & & & & {{\scriptsize {\em Last but not least, MDD is dangerous because innovation distracts. }}}  \\

    \cline{4-5}

 & & & \multirow{5}[1]{*}{\parbox{2.5cm}{Tools as a way to increase complexity}} & {{\scriptsize {\em Most existing modeling and model-driven tools don’t include a full featured versioning system. That’s a pain when working with large teams. }}}  \\

& & & &  \\
& & & &  \\

   \hline

 \multirow{7}[1]{*}{\cite{Cabot:2009:Online}} & \multirow{7}[1]{*}{2009} & \multirow{7}[1]{*}{\parbox{2cm}{Web site}} & \multirow{8}[1]{*}{\parbox{2.5cm}{Implicit questions derived from the MDE adoption itself}} &  \multicolumn{1}{p{10 cm}|}{{\scriptsize {\em Stephen is confident we can make it to the other side because after a “peak of inflated expectations” and, its result, the {\em trough of disillusionment} (where UML/modeling become unfashionable due to failed expectations) we are now progressing through the {\em Slope of enlightment} in the technology hype cycle , where software engineers start to understand how to best use modeling and its potential benefits .}}}  \\

 & & & & {{\scriptsize {\em However, he pointed out several hurdles that still impede our progress, mainly related to our lack of ability to prove to practitioners that modeling could work for them and improve their productivity.}}}  \\

   \hline

    \end{tabular}%
  \caption{Industrial issues evidenced (20/38).}
  \label{Table20}%
\end{table}%


\clearpage
\newpage

\begin{table}[htbp]
  \centering
    \begin{tabular}{|p{2 cm}|p{1 cm}|p{2 cm}|p{3 cm}|p{9 cm}|}
      \hline
    \multicolumn{1}{|c|}{{\bf Source}} & \multicolumn{1}{|c|}{{\bf  Year}}  & \multicolumn{1}{|c|}{{\bf Type}} & \multicolumn{1}{|c|}{{\bf Issue}} & \multicolumn{1}{|c|}{{\bf Statements}} \\
    \hline

 \multirow{10}[1]{*}{\cite{Pierson:2007:Online}} & \multirow{10}[1]{*}{2007} & \multirow{10}[1]{*}{\parbox{2cm}{Web site}} & \multirow{7}[1]{*}{\parbox{2.5cm}{Implicit questions derived from the MDE adoption itself}} &  \multicolumn{1}{p{9 cm}|}{{\scriptsize {\em For me, the bigger issue—at least, for [the] time being—is whether or not code will remain first-class artifacts throughout the software life cycle, or will the model-driven development technology [that] we bring to market now make the same mistake as CASE, trying to replace code entirely by models? That may be possible, on some distant day in the future; today, it's just not practical}}}  \\

 & & & & \\
 & & & & \\

    \cline{4-5}

 & & & \multirow{5}[1]{*}{\parbox{2.5cm}{Tools as a way to increase complexity}} & {{\scriptsize {\em If getting modeling tools support for new modeling language requires significance amount of coding, off soon come at the point when the amount of tool code changes a new modeling language feature would require make that change impractical. That's the point with which modeling language becomes to go the way [of] the dinosaurs.}}}  \\

\hline

 \multirow{10}[1]{*}{\cite{Bambrilla:2016:Online}} & \multirow{10}[1]{*}{2016} & \multirow{10}[1]{*}{\parbox{2cm}{Web site}} & \multirow{7}[1]{*}{\parbox{2.5cm}{Implicit questions derived from the MDE adoption itself}} &  \multicolumn{1}{p{9 cm}|}{{\scriptsize {\em Is the real problem of MDE about the word {\bf modeling}? In any other fields modelling is implicit and obvious - Bran Selic}}}  \\

 & & & & {{\scriptsize {\em The flow of model must be clarified: traceability, refinement, model integration are crucial. You must grant syntactic and semantic coherence.}}}  \\
 & & & & \\
 & & & & \\
    \cline{4-5}

 & & & \multirow{5}[1]{*}{\parbox{2.5cm}{Tools as a way to increase complexity}} & {{\scriptsize {\em You also need a coherent infrastructure of tools and artefacts, that grants logic integration. You cannot obtain coherence of models without coherence of tools.}}}  \\

 & & & & {{\scriptsize {\em You need a lot of automation, otherwise you won't get practical maturity.}}}  \\

\hline

    \end{tabular}%
  \caption{Industrial issues evidenced (21/38).}
  \label{Table21}%
\end{table}%


\clearpage
\newpage

\begin{table}[htbp]
  \centering
    \begin{tabular}{|p{2 cm}|p{1 cm}|p{2 cm}|p{3 cm}|p{10 cm}|}
      \hline
    \multicolumn{1}{|c|}{{\bf Source}} & \multicolumn{1}{|c|}{{\bf  Year}}  & \multicolumn{1}{|c|}{{\bf Type}} & \multicolumn{1}{|c|}{{\bf Issue}} & \multicolumn{1}{|c|}{{\bf Statements}} \\
    \hline

 \multirow{20}[1]{*}{\cite{DenHaan:2011(2):Online}} & \multirow{20}[1]{*}{2012} & \multirow{20}[1]{*}{\parbox{2cm}{Web site}} & \multirow{20}[1]{*}{\parbox{2.5cm}{Tools as a way to increase complexity}} &  \multicolumn{1}{p{10 cm}|}{{\scriptsize {\em Earning curve - modeling tools have been evolving so rapidly, that I'm hard pressed to find engineers that deeply understand the tool. I still find you are only as good as your modeling tool.}}}  \\

 & & & & {{\scriptsize {\em  Too structured - Personally, I've been in situations where I found that the modeling tool was simply too structured to let me describe everything I needed to describe.}}}  \\
 & & & & {{\scriptsize {\em  Cost of tuning the tool - every time I've tried to autogenerate code, I've ended up manually reworking the code once I see what the tool thought was right.}}}  \\
 & & & & {{\scriptsize {\em  Time to market - I've experienced definite problems when in a situation where the need for working software was urgent. If the project and team are small enough, I see no reason to waste time on a modeling tool when the time can be spent coding and testing.}}}  \\
 & & & & {{\scriptsize {\em  Cost of failure - when I've seen projects run away from modeling tools, it's because of the high cost of failure - to use the tools, you need every developer to be involved. That's a big investment in training and hands on learning, and a very costly mistake if someone has set up the model badly.}}}  \\
 & & & & {{\scriptsize {\em  Missing Refactoring Support: Lets guess I want to model the entities of my datamodel with MDA (Typical usecase No. 1). If I have my model in, lets say, an UML diagram, and I change it, nothing of my code changes with it (at least the generated classes), and instead of having still a working app with better named attributes, I get a lot of errors I have to correct manually.}}}  \\
 & & & & {{\scriptsize {\em  Missing debugging support: Usually translations from model to code are done by having some transformation language at hand. This would be no problem usually, but when we debug, we optimally should not be worrying of the code we generate, and a debugger should step into the transformation model.}}}  \\
 & & & & {{\scriptsize {\em  Transformations are hard to test: If you use transformations in a specialized IDE, they are done by the IDEs compiler .}}}  \\

\hline

    \end{tabular}%
  \caption{Industrial issues evidenced (22/38).}
  \label{Table22}%
\end{table}%


\clearpage
\newpage

\begin{table}[htbp]
  \centering
    \begin{tabular}{|p{2 cm}|p{1 cm}|p{2 cm}|p{3 cm}|p{10 cm}|}
      \hline
    \multicolumn{1}{|c|}{{\bf Source}} & \multicolumn{1}{|c|}{{\bf  Year}}  & \multicolumn{1}{|c|}{{\bf Type}} & \multicolumn{1}{|c|}{{\bf Issue}} & \multicolumn{1}{|c|}{{\bf Statements}} \\
    \hline

 \multirow{10}[1]{*}{\cite{Monkey:2016:Online}} & \multirow{10}[1]{*}{2015} & \multirow{10}[1]{*}{\parbox{2cm}{Web site (Survey) }} & \multirow{10}[1]{*}{\parbox{2.5cm}{Implicit questions derived from the MDE adoption itself}} &  \multicolumn{1}{p{10 cm}|}{{\scriptsize {\em We would like your opinion on how the use of different MDE activities on projects affects productivity and maintainability.}}}  \\

 & & & & {{\scriptsize {\em  ¿Does using MDE require you to carry out significant extra training in modeling?.}}}  \\
 & & & & {{\scriptsize {\em Is integrating generated code into your existing projects a significant problem?.}}}  \\
 & & & & {{\scriptsize {\em Does MDE prevent you from responding to business opportunities?.}}}  \\
 & & & & {{\scriptsize {\em Does model-based testing require significant additional time to specify the models at a sufficient level of detail for testing?.}}}  \\
 & & & & {{\scriptsize {\em Are MDE tools too expensive?.}}}  \\

\hline

 \multirow{18}[1]{*}{\cite{Platania:2016:Online}} & \multirow{18}[1]{*}{2016} & \multirow{18}[1]{*}{\parbox{2cm}{Web site}} & \multirow{10}[1]{*}{\parbox{2.5cm}{MDA is not enough}} &  \multicolumn{1}{p{10 cm}|}{{\scriptsize {\em But more and more what I get as a feedback is that MDA is dead, it's don't work, it's passe'.}}}  \\

 & & & & {{\scriptsize {\em  Giuseppe, another point: you don't need use OMG's MDA to put MDD in practice. I think a simple and easy approach is enough. You have to be simple to be followed by the people. If you get discuss the things in therms of diagram, generated code and aware the codders the importance to respect the original design, you already is working in MDD. Maybe, you won't even need jump to MDE or MDA at all.}}}  \\

 & & & & {{\scriptsize {\em MDA is very fragile.}}}  \\
\cline{4-5}

 & & & \multirow{7}[1]{*}{\parbox{2.5cm}{Implicit questions derived from the MDE adoption itself}} & {{\scriptsize {\em What the industry needs are the methods that survive inexperienced stuff, burning deadlines, budget cuts and 20+ years in maintenance with several generations of developers.}}}  \\
 & & & & {{\scriptsize {\em Agile is about project management, MBSE is about systems engineering and the reuse of models. Problems begin when agile is applied indiscriminately to every kind of project, and MBSE is understood as code generation.}}}  \\

 & & & &   \\

\hline

    \end{tabular}%
  \caption{Industrial issues evidenced (23/38).}
  \label{Table23}%
\end{table}%


\clearpage
\newpage

\begin{table}[htbp]
  \centering
    \begin{tabular}{|p{2 cm}|p{1 cm}|p{2 cm}|p{3 cm}|p{10 cm}|}
      \hline
    \multicolumn{1}{|c|}{{\bf Source}} & \multicolumn{1}{|c|}{{\bf  Year}}  & \multicolumn{1}{|c|}{{\bf Type}} & \multicolumn{1}{|c|}{{\bf Issue}} & \multicolumn{1}{|c|}{{\bf Statements}} \\
    \hline

 \multirow{10}[1]{*}{\cite{clark2012exploiting}} & \multirow{10}[1]{*}{2012} & \multirow{10}[1]{*}{\parbox{2cm}{Journal Paper}} & \multirow{10}[1]{*}{\parbox{2.5cm}{Implicit questions derived from the MDE adoption itself }} &  \multicolumn{1}{p{10 cm}|}{{\scriptsize {\em The grand challenge of replacing programming by modeling has not been achieved. }}}  \\

 & & & & {{\scriptsize {\em UML Usage The current state-of-the-art in industrial MDD ranges from informal
sketch-based modelling to the generation of code skeletons from blueprints. Practitioners are often sceptical about UML CASE tools and their ability to generate complete applications. It is not currently realistic for a small company to make grand claims about the ability of MDD to deliver large-scale business value in the general case.}}}  \\

 & & & \multirow{5}[1]{*}{\parbox{2.5cm}{\&}}  & {{\scriptsize {\em Investor Confidence Investors need to know that there is a significant business that will arise from the use of MDD. Industry scepticism and the lack of high-profile business cases makes it difficult for MDD startups to attract seed funding.}}}  \\

 & & & \multirow{5}[1]{*}{\parbox{2.5cm}{MDA is not enough}} & {{\scriptsize {\em Novelty MDD approaches are still considered novel and this raises a significant barrier for adoption by large companies. In most situations companies will trust the technologies that they are familiar with.}}}  \\

 & & & & {{\scriptsize {\em Model management is a problem for largescale MDD adoption where multi-developer distributed projects are a requirement. Both companies described in this article undertook a major rewrite of their technology. This is typical of advanced technology platforms and is very difficult to achieve success-  fully where existing customers have large model repositories.}}}  \\

 & & & & {{\scriptsize {\em In our experience, some industrial uses of MDD are badly performed and lead to misjudged opinions. This is perhaps related to the immaturity of MDD and the high-levels of technical expertise required to use tools.}}}  \\
 
\hline

    \end{tabular}%
  \caption{Industrial issues evidenced (24/38).}
  \label{Table24}%
\end{table}%


\clearpage
\newpage

\begin{table}[htbp]
  \centering
    \begin{tabular}{|p{2 cm}|p{1 cm}|p{2 cm}|p{3 cm}|p{10 cm}|}
      \hline
    \multicolumn{1}{|c|}{{\bf Source}} & \multicolumn{1}{|c|}{{\bf  Year}}  & \multicolumn{1}{|c|}{{\bf Type}} & \multicolumn{1}{|c|}{{\bf Issue}} & \multicolumn{1}{|c|}{{\bf Statements}} \\
    \hline

 \multirow{10}[1]{*}{\cite{clark2012exploiting}} & \multirow{10}[1]{*}{2012} & \multirow{10}[1]{*}{\parbox{2cm}{Journal Paper}} & \multirow{10}[1]{*}{\parbox{2.5cm}{Implicit questions derived from the MDE adoption itself}} &  \multicolumn{1}{p{10 cm}|}{{\scriptsize {\em Even when generating programs from models, the temptation to edit the resulting code is overwhelming. This places MDD tools that are unable to cope with the scale of the changes, in an unfavourable light.}}}  \\

 & & & & {{\scriptsize {\em Practitioners often argue that the risk associated with generating code from platform independent models is too great and of limited practical value. Where consensus is reached, it is often the case that PIMs include many aspects of the target platforms in order to support code generation.}}}  \\

 & & & \multirow{5}[1]{*}{\parbox{2.5cm}{\&}} & {{\scriptsize {\em  Relational vs OO Database administrators will often complain about the impedance mismatch between a relational and an OO model. }}}  \\

 & & & & {{\scriptsize {\em Once multiple models and associated code have been generated there is often a maintenance problem that is cited by companies as a reason for not adopting MDD. Multiple versions of UML serialization formats and a general lack of viable interoperability between MDD tools is cited as a business risk.}}}  \\

 & & & \multirow{1}[1]{*}{\parbox{2.5cm}{MDA is not enough}} & {{\scriptsize {\em MDD is often seen by industry as too heavyweight and complex. Problems arising from early adoption can often be perceived to exist long after technologies have matured.}}}  \\

\hline

    \end{tabular}%
  \caption{Industrial issues evidenced (25/38).}
  \label{Table25}%
\end{table}%


\clearpage
\newpage

\begin{table}[htbp]
  \centering
    \begin{tabular}{|p{2 cm}|p{1 cm}|p{2 cm}|p{3 cm}|p{10 cm}|}
      \hline
    \multicolumn{1}{|c|}{{\bf Source}} & \multicolumn{1}{|c|}{{\bf  Year}}  & \multicolumn{1}{|c|}{{\bf Type}} & \multicolumn{1}{|c|}{{\bf Issue}} & \multicolumn{1}{|c|}{{\bf Statements}} \\
    \hline

 \multirow{10}[1]{*}{\cite{mohagheghi2008mde}} & \multirow{10}[1]{*}{2008} & \multirow{10}[1]{*}{\parbox{2cm}{Workshop Paper}} & \multirow{10}[1]{*}{\parbox{2.5cm}{Implicit questions derived from the MDE adoption itself}} &  \multicolumn{1}{p{10 cm}|}{{\scriptsize {\em First of all, the UML standard has evolved but, with this evolution, the syntax has become even more complex and the necessary supporting mechanisms and tools for dealing with this added complexity are not yet available. Even something as conceptually simple as exporting a UML diagram from one tool to another has not been accomplished yet with ease. On the other hand, developing a DSM solution requires high skills related to meta-modeling and tool development.}}}  \\

 & & & & {{\scriptsize {\em Also a big concern with Domain-Specific Languages (DSLs) is getting the people in that domain to agree upon a standard
syntax. Another challenge is having that DSL interact properly with anything outside of its domain, having a different underlying syntax to that of other languages.}}}  \\

 & & & & {{\scriptsize {\em Another challenge for organizations wanting to get started in MDE, closely related with the previous idea of managing all these artifacts, is that they may end up dealing with more complexity than anticipated at first. From our experience in the field we have gotten the impression that, if not adequately managed, the development of complex systems with MDE gets treated with more complexity. }}}  \\

\cline{4-5}

 & & & \multirow{5}[1]{*}{\parbox{2.5cm}{MDA is not enough}} & {{\scriptsize {\em }}}  \\

 & & & & \\

 & & & \multirow{5}[1]{*}{\parbox{2.5cm}{\&}} & {{\scriptsize {\em  Arlier efforts in modeling failed due to the complexity of UML, the lack of proper tools and the inability to maintain models in synch with code, among other issues. Due to the above problems with UML, we decided to develop our own programming tools and frameworks addressing the problem domain. }}}  \\

 & & & \multirow{1}[1]{*}{\parbox{2.5cm}{Tools as a way to increase complexity}} & {{\scriptsize {\em }}}  \\

 & & & & \\
 & & & & \\
 & & & & \\
\hline

    \end{tabular}%
  \caption{Industrial issues evidenced (26/38).}
  \label{Table26}%
\end{table}%


\clearpage
\newpage

\begin{table}[htbp]
  \centering
    \begin{tabular}{|p{2 cm}|p{1 cm}|p{2 cm}|p{3 cm}|p{10 cm}|}
      \hline
    \multicolumn{1}{|c|}{{\bf Source}} & \multicolumn{1}{|c|}{{\bf  Year}}  & \multicolumn{1}{|c|}{{\bf Type}} & \multicolumn{1}{|c|}{{\bf Issue}} & \multicolumn{1}{|c|}{{\bf Statements}} \\
    \hline

 \multirow{10}[1]{*}{\cite{whittle2015taxonomy}} & \multirow{10}[1]{*}{2015} & \multirow{10}[1]{*}{\parbox{2cm}{Journal Paper}} & \multirow{7}[1]{*}{\parbox{2.5cm}{Implicit questions derived from the MDE adoption itself}} &  \multicolumn{1}{p{10 cm}|}{{\scriptsize {\em MDE can be very effective, but it takes effort to make it work. The majority of our interviewees were very successful with MDE but all of them either built their own modeling tools, made heavy adaptations of off-the-shelf tools, or spent a lot of time findingways towork around tools.}}}  \\

 & & & & \\
 & & & & \\
 & & & & \\

\cline{4-5}

 & & & \multirow{10}[1]{*}{\parbox{2.5cm}{Tools as a way to increae complexity}} &  \multicolumn{1}{p{10 cm}|}{{\scriptsize {\em Indeed, this suggests that current tools are a barrier to success rather than an enabler and  ``the fact that people are struggling with the tools...and succeed nonetheless requires a certain level of enthusiasm and competence".}}}  \\

 & & &  &  \multicolumn{1}{p{10 cm}|}{{\scriptsize {\em Our interviewees emphasized tool immaturity, complexity, and lack of usability as major barriers.}}}  \\

 & & &  &  \multicolumn{1}{p{10 cm}|}{{\scriptsize {\em A curious paradox of MDE is that it was developed as a way to improve portability [18]. However, time and again issues of migration and versioning came up in our interviews: ``[XX] have burned a lot of money to build their own tool which they stopped doing because they lost theirmodelswhen the [YY] version changed."}}}  \\

 & & &  &  \multicolumn{1}{p{10 cm}|}{{\scriptsize {\em Unfortunately, there is also a clear gap in the way that vendors market their tools and their real capabilities in terms of this low-level approach. A.s a result, many MDE applications fail due to expectations that have not been managed properly.}}}  \\

 & & & & {{\scriptsize {\em In other cases, interviewees reported that MDE tools can make certification more difficult as current government certification processes are not set up to deal with auto-generated code.}}}  \\

\hline

    \end{tabular}%
  \caption{Industrial issues evidenced (27/38).}
  \label{Table27}%
\end{table}%


\clearpage
\newpage

\begin{table}[htbp]
  \centering
    \begin{tabular}{|p{2 cm}|p{1 cm}|p{2 cm}|p{3 cm}|p{10 cm}|}
      \hline
    \multicolumn{1}{|c|}{{\bf Source}} & \multicolumn{1}{|c|}{{\bf  Year}}  & \multicolumn{1}{|c|}{{\bf Type}} & \multicolumn{1}{|c|}{{\bf Issue}} & \multicolumn{1}{|c|}{{\bf Statements}} \\
    \hline

 \multirow{10}[1]{*}{\cite{whittle2015taxonomy}} & \multirow{10}[1]{*}{2015} & \multirow{10}[1]{*}{\parbox{2cm}{Journal Paper}}  & \multirow{7}[1]{*}{\parbox{2.5cm}{Implicit questions derived from the MDE adoption itself}} & {{\scriptsize {\em }}}  \\

 & & & & \\
 & & & & \\
 & & & & \\
 & & & & \\

 & & & \multirow{5}[1]{*}{\parbox{2.5cm}{\&}} & {{\scriptsize {\em It is ironic that MDE was introduced to help deal with the essential complexity of systems, but in many cases, adds accidental complexity.}}}  \\
 & & & & \\
 & & & & \\

 & & & \multirow{1}[1]{*}{\parbox{2.5cm}{Tools as a way to increase complexity}} & {{\scriptsize {\em }}}  \\
 & & & & \\
 & & & & \\
 & & & & \\
\hline

    \end{tabular}%
  \caption{Industrial issues evidenced (28/38).}
  \label{Table28}%
\end{table}%


\clearpage
\newpage

\begin{table}[htbp]
  \centering
    \begin{tabular}{|p{2 cm}|p{1 cm}|p{2 cm}|p{3 cm}|p{10 cm}|}
      \hline
    \multicolumn{1}{|c|}{{\bf Source}} & \multicolumn{1}{|c|}{{\bf  Year}}  & \multicolumn{1}{|c|}{{\bf Type}} & \multicolumn{1}{|c|}{{\bf Issue}} & \multicolumn{1}{|c|}{{\bf Statements}} \\
    \hline

 \multirow{20}[1]{*}{\cite{marin2014key}} & \multirow{20}[1]{*}{2014} & \multirow{20}[1]{*}{\parbox{2cm}{Conference Paper}}  & \multirow{20}[1]{*}{\parbox{2.5cm}{Tools as a way to increase complexity}} & {{\scriptsize {\em Nowadays, more than ever, it is necessary to provide user interfaces that take into account quality characteristics such as usability (ISO/IEC, 2001); and new characteristics of hardware devices, such as touch panels in desktop and laptop computers.}}}  \\

 & & & & {{\scriptsize {\em 100\% of subjects response in the exploratory study that the MDD tools must provide verification mechanisms.}}}  \\
 & & & & {{\scriptsize {\em MDD tools must provide model-based testing approaches that will be focus in holistic models.}}}  \\
 & & & & {{\scriptsize {\em MDD tools should allow the execution of models even though they are incomplete (but valid). The main idea is not to wait until the model is finished to see how it looks like the software obtained from the part of the model that is already specified. }}}  \\
 & & & & {{\scriptsize {\em A suitable MDD tool must offer a number of predefined transformations for assuring a complete model transformation.}}}  \\
 & & & & {{\scriptsize {\em MDD tools should significantly reduce time and efforts, and simplify the development of final software products. }}}  \\
 & & & & {{\scriptsize {\em MDD tools work with platform independent models; the tool must support the transformation to executable code not only to a variety of languages, but also to different architecture design patterns. }}}  \\
 & & & & {{\scriptsize {\em MDD tools offer verification of semantic defects in order to prevent faults when the generated system is executed.}}}  \\
 & & & & {{\scriptsize {\em  MDD tools do not offer options to generate tests or simulation artifacts in order to properly validate the models, rather than testing the code once the system is generated.}}}  \\
 & & & & {{\scriptsize {\em Regarding the transformation, even though the tools analyzed offer refactoring and reverse engineering features, they do not provide facilities to customize transformations in particular situations.}}}  \\
 & & & & {{\scriptsize {\em With respect to efficiency and architecture, commercial tools such as Rational or Integranova can export to a number of different architectures, but at open-source tools this option is available}}}  \\

\hline

    \end{tabular}%
  \caption{Industrial issues evidenced (29/38).}
  \label{Table29}%
\end{table}%


\clearpage
\newpage

\begin{table}[htbp]
  \centering
    \begin{tabular}{|p{2 cm}|p{1 cm}|p{2 cm}|p{3 cm}|p{10 cm}|}
      \hline
    \multicolumn{1}{|c|}{{\bf Source}} & \multicolumn{1}{|c|}{{\bf  Year}}  & \multicolumn{1}{|c|}{{\bf Type}} & \multicolumn{1}{|c|}{{\bf Issue}} & \multicolumn{1}{|c|}{{\bf Statements}} \\
    \hline

 \multirow{20}[1]{*}{\cite{Aranda2012}} & \multirow{20}[1]{*}{2012} & \multirow{20}[1]{*}{\parbox{2cm}{Conference Paper}}  & \multirow{7}[1]{*}{\parbox{2.5cm}{Implicit questions derived from the MDE adoption itself}} & {{\scriptsize {\em First, MDE succeeds in bringing software development closer to the subject matter experts, but an important (if at times menial) subset of software development activities still needs to be performed by people other than the subject matter experts—people with significant software development skills.}}}  \\

 & & & & {{\scriptsize {\em MacDonald et al., who claim that MDE does not lead to an improvement in efficiency, effectiveness, or productivity—at least not in the context of projects with a large amount of legacy code.}}}  \\

\cline{4-5}

 & & &  \multirow{10}[1]{*}{\parbox{2.5cm}{Organizational support for the MDE adoption}} & {{\scriptsize {\em Switching to MDE may disrupt the organizational structure and alter its balance, which creates morale and power problems that transitioning groups should consider.}}}  \\
 & & & & {{\scriptsize {\em MDE represents a migration to an underpopulated cultural and institutional landscape. The tools, training, and expectations of professionals under MDE are not as well developed and established as those under more traditional software development dynamics.}}}  \\
 & & & & {{\scriptsize {\em At the outset, it is unclear whether MDE can be one such revolutionary technology. On one hand, MDE could upset the whole communication and coordination structure, bringing many roles into obsolescence, and eliminating the need for time-consuming and inefficient structures.}}}  \\

\cline{4-5}

 & & & \multirow{5}[1]{*}{\parbox{2.5cm}{Organizational support for the MDE adoption}} & {{\scriptsize {\em While MDE has matured over the years to the point where it can sustain the development of products of critical importance and of high quality, as GM’s automotive software needs to be, its institutional infrastructure is still underdeveloped, and transitioning practitioners will find that, both technically and organizationally, many things they took for granted need to be built again.}}}  \\

 & & & \multirow{1}[1]{*}{\parbox{2.5cm}{\&}} & {{\scriptsize {\em  This introduction of MDE tools, however, brought a disruption in the work arrangement.}}}  \\
 & & & \multirow{1}[1]{*}{\parbox{2.5cm}{Tools as a way to increase complexity}} & {{\scriptsize {\em}}}  \\
 & & & & \\
 & & & & \\
 & & & & \\

\hline

    \end{tabular}%
  \caption{Industrial issues evidenced (30/38).}
  \label{Table30}%
\end{table}%


\clearpage
\newpage

\begin{table}[htbp]
  \centering
    \begin{tabular}{|p{2 cm}|p{1 cm}|p{2 cm}|p{3 cm}|p{10 cm}|}
      \hline
    \multicolumn{1}{|c|}{{\bf Source}} & \multicolumn{1}{|c|}{{\bf  Year}}  & \multicolumn{1}{|c|}{{\bf Type}} & \multicolumn{1}{|c|}{{\bf Issue}} & \multicolumn{1}{|c|}{{\bf Statements}} \\
    \hline

 \multirow{20}[1]{*}{\cite{baker2005model}} & \multirow{20}[1]{*}{2005} & \multirow{20}[1]{*}{\parbox{2cm}{Conference Paper}}  & \multirow{20}[1]{*}{\parbox{2.5cm}{Implicit questions derived from the MDE adoption itself}} & {{\scriptsize {\em First, MDE succeeds in bringing software development closer to the subject matter experts, but an important (if at times menial) subset of software development activities still needs to be performed by people other than the subject matter experts—people with significant software development skills.}}}  \\

 & & & & {{\scriptsize {\em Lack of rigorous models for test generation. Even though it is common for system architects and designers to use scenario-based notations, they typically do not contain the rigor needed for machine processing. We also found that architects and designers were reluctant to invest the extra effort needed to develop rigorous models since the benefit of automated test generation did not immediately justify the extra effort within their project scope.}}}  \\

 & & & & {{\scriptsize {\em Lack of Abstraction. Platform Specifics: Often we find that system architects and designers develop requirements that contain either implicit or explicit assumptions about the implementation.}}}  \\

 & & & & {{\scriptsize {\em Incompleteness: Typically requirements are defined either through use cases or through scenario-based models. }}}  \\
 & & & & {{\scriptsize {\em Quality: Determining the adequacy of requirements is a common concern.}}}  \\

 & & & & {{\scriptsize {\em Lack of Well Defined Semantics. Motorola projects have encountered issues with language semantics with virtually every modeling language used. For example, SA/SD had no defined process language and advocated natural language, tool vendors have added their own language extensions, and even UML 2.0 contains semantic variation points.}}}  \\

 & & & & {{\scriptsize {\em We have observed that the current state of corporate MDE usage is characterized by isolated models. For example, even in a highly coupled system such as iDEN, each model exists separately. That is, they interact with each other only in the target network and are neither currently being modeled as a whole nor being leveraged for simulation and consistency checking as a whole.}}}  \\

\hline

    \end{tabular}%
  \caption{Industrial issues evidenced (31/38).}
  \label{Table31}%
\end{table}%


\clearpage
\newpage

\begin{table}[htbp]
  \centering
    \begin{tabular}{|p{2 cm}|p{1 cm}|p{2 cm}|p{3 cm}|p{10 cm}|}
      \hline
    \multicolumn{1}{|c|}{{\bf Source}} & \multicolumn{1}{|c|}{{\bf  Year}}  & \multicolumn{1}{|c|}{{\bf Type}} & \multicolumn{1}{|c|}{{\bf Issue}} & \multicolumn{1}{|c|}{{\bf Statements}} \\
    \hline

 \multirow{20}[1]{*}{\cite{baker2005model}} & \multirow{20}[1]{*}{2005} & \multirow{20}[1]{*}{\parbox{2cm}{Conference Paper}}  & \multirow{6}[1]{*}{\parbox{2.5cm}{Organizational support for the MDE adoption}} & {{\scriptsize {\em We have observed that many teams encounter major obstacles in adopting MDE due to the lack of a well defined MDE process, missing skill sets, and inflexibility in changing the existing culture. Without a well defined MDE process, teams that adopt MDE tend to use a ``trial and error" approach and encounter the same set of pitfalls others have already experienced.}}}  \\

\cline{4-5}

 & & & \multirow{10}[1]{*}{\parbox{2.5cm}{Tools as a way to increase complexity}} & {{\scriptsize {\em Lack of Common Tools. The ready availability of third-party and internal tools for modeling and code generation has led to a wide diversity of processes, languages, etc. Even within a “single” language such as UML, there are several issues such as the inability to completely transfer models between tools, use of vendor-specific extensions, lack of complete UML support, and code generation support for different subsets of UML.}}}  \\

 & & &  & {{\scriptsize {\em Third-party MDE tools often do not scale well to the sizes needed for modeling   real telecommunications systems. We have encountered issues with the ability of tools to load, save, compare, and generate code from large models.}}}  \\

 & & &  & {{\scriptsize {\em No single tool supports a comprehensive MDE environment, allowing full use of current tools and processes. To this end, integration of modeling concepts and tools becomes problematic.}}}  \\

\hline

    \end{tabular}%
  \caption{Industrial issues evidenced (32/38).}
  \label{Table32}%
\end{table}%


\clearpage
\newpage

\begin{table}[htbp]
  \centering
    \begin{tabular}{|p{2 cm}|p{1 cm}|p{2 cm}|p{3 cm}|p{10 cm}|}
      \hline
    \multicolumn{1}{|c|}{{\bf Source}} & \multicolumn{1}{|c|}{{\bf  Year}}  & \multicolumn{1}{|c|}{{\bf Type}} & \multicolumn{1}{|c|}{{\bf Issue}} & \multicolumn{1}{|c|}{{\bf Statements}} \\
    \hline

 \multirow{20}[1]{*}{\cite{Hoang:2012:Online}} & \multirow{20}[1]{*}{2012} & \multirow{20}[1]{*}{\parbox{2cm}{Web Page}}  & \multirow{15}[1]{*}{\parbox{2.5cm}{Tools as a way to increase complexity}} & {{\scriptsize {\em The UML part was more difficult to apply, because the limitation of tools: Manually synchronization between artifacts: If your source code is changed, you have to manually update all the related artifacts (from requirement to deployments documents) to fit with the current reality.}}}  \\

 & & &  & {{\scriptsize {\em Current MDD tools do not allow customization of user interaction (issue of e.g. excessive mouse clicks, but also manipulation of modeling elements). In my experience, user interaction is also very-much domain specific.}}}  \\

 & & &  & {{\scriptsize {\em I would like to stress that today even the best DSM tools (aka language workbenches) that allow efficient customization of modeling concepts have a hardcoded "model" of user interaction with model.}}}  \\

 & & &  & {{\scriptsize {\em Do that to the full extent requires each modeling tool to be built from scratch, with its own requirements analysis, design, hand-coded implementation, testing etc. Obviously, for all but the largest numbers of users and most static set of requirements, it makes more sense to accept some generic tool functionality, in order to get a modeling tool faster.}}}  \\

\cline{4-5}

 & & & \multirow{6}[1]{*}{\parbox{2.5cm}{Implicit questions derived from the MDE adoption itself}} & {{\scriptsize {\em The transformation rules and the domain specific languages (like UML, BPMN) are unnecessary complicated. People can use the notations in different ways, and they have to invest a significant time to learn in order to use in a reasonable way.}}}  \\

 & & &  &   \\
 & & &  &   \\

\hline

    \end{tabular}%
  \caption{Industrial issues evidenced (33/38).}
  \label{Table33}%
\end{table}%


\clearpage
\newpage

\begin{table}[htbp]
  \centering
    \begin{tabular}{|p{2 cm}|p{1 cm}|p{2 cm}|p{3 cm}|p{10 cm}|}
      \hline
    \multicolumn{1}{|c|}{{\bf Source}} & \multicolumn{1}{|c|}{{\bf  Year}}  & \multicolumn{1}{|c|}{{\bf Type}} & \multicolumn{1}{|c|}{{\bf Issue}} & \multicolumn{1}{|c|}{{\bf Statements}} \\
    \hline

    \multirow{7}[6]{*}{{\cite{DenHaan:2011:Online}}} & \multirow{7}[6]{*}{2011} & \multirow{7}[6]{*}{\parbox{2cm}{Web Site}} & \multirow{5}[1]{*}{\parbox{3cm}{Implicit questions derived from the MDE  adoption itself }} &  \multicolumn{1}{p{10 cm}|}{{\scriptsize {\em They idea of MDD is to limit the flexibility in favor of simplicity and productivity. Hence, it isn’t possible to create everything you want.}}} \\

  &       &       &       & \\ 
  &       &       &       & \\ 
  &       &       &       & \\ 

\cline{4-5}

          &       &       &    \multirow{5}[1]{*}{\parbox{3cm}{Tools as a way to increase complexity}}    & \multicolumn{1}{p{10 cm}|}{{\scriptsize {\em The problem however is: if your MDD tool reaches a certain level, development isn’t the slowest part of developing software anymore, deploying and taking it into production is. MDD can’t help you out here.}}}  \\

  &       &       &       & \\ 

\hline

    \multirow{7}[6]{*}{{\cite{whittle2013industrial}}} & \multirow{7}[6]{*}{2013} & \multirow{7}[6]{*}{\parbox{2cm}{Conference Paper}} & \multirow{10}[1]{*}{\parbox{3cm}{Implicit questions derived from the MDE  adoption itself }} &  \multicolumn{1}{p{10 cm}|}{{\scriptsize {\em It is ironic that MDE was introduced to help deal with the essential complexity of systems, but in many cases, adds accidental complexity. }}} \\

  &       &       &       & \multicolumn{1}{p{10 cm}|}{{\scriptsize {\em  ``I don’t think you gain advantage in solving all kinds of problems in modeling.” There is a danger of over-engineering the solution:  You would try to do some smart modeling, or stuff and you would fail. After a while you would end up in a worse place than if you had done this in C++".}}} \\

  &       &       &       & \multicolumn{1}{p{10 cm}|}{{\scriptsize {\em The big conclusion of our studies is that MDE can work, but it is a struggle. MDE tools do not seem to support those who try. We need simpler tools and more focus on the underlying processes. MDE tools also need to be more resilient: as with any new method, MDE is highly dependent on a range of technical, social and organizational factors. }}} \\

\hline

    \end{tabular}%
  \caption{Industrial issues evidenced (34/38).}
  \label{Table34}%
\end{table}%


\clearpage
\newpage

\begin{table}[htbp]
  \centering
    \begin{tabular}{|p{2 cm}|p{1 cm}|p{2 cm}|p{3 cm}|p{10 cm}|}
      \hline
    \multicolumn{1}{|c|}{{\bf Source}} & \multicolumn{1}{|c|}{{\bf  Year}}  & \multicolumn{1}{|c|}{{\bf Type}} & \multicolumn{1}{|c|}{{\bf Issue}} & \multicolumn{1}{|c|}{{\bf Statements}} \\
    \hline

    \multirow{15}[6]{*}{{\cite{whittle2013industrial}}} & \multirow{15}[6]{*}{2013} & \multirow{15}[6]{*}{\parbox{2cm}{Conference Paper}} & \multirow{15}[1]{*}{\parbox{3cm}{Tools as a way to increase complexity}} &  \multicolumn{1}{p{10 cm}|}{{\scriptsize {\em The only accounts of easy-to-use, intuitive tools came from those who had developed tools themselves for bespoke purposes. Indeed, this suggests that current tools are a barrier to success rather than an enabler and ``the fact that people are struggling with the tools. . . and succeed nonetheless requires a certain level of enthusiasm and competence."}}} \\

  &       &       &       & \multicolumn{1}{p{10 cm}|}{{\scriptsize {\em  Our interviewees emphasized tool immaturity, complexity and lack of usability as major barriers.  More generally, tools are often very powerful, but it is too difficult for users to access that power; or, in some cases, they do not really need that power and require something much simpler: ``I was really impressed with the power of it and on the other hand I saw windows popping up everywhere. . . at the end I thought I still really have no idea how to use this tool and I have only seen a glimpse of the power that it has".}}} \\

  &       &       &       & \multicolumn{1}{p{10 cm}|}{{\scriptsize {\em  A curious paradox of MDE is that it was developed as a way to improve portability ... However, time and again issues of migration and versioning came up in our interviews: ``[XX] have burned a lot of money to build their own tool which they stopped doing because they lost their models when the [YY] version changed".}}} \\

  &       &       &       & \multicolumn{1}{p{10 cm}|}{{\scriptsize {\em  The cost of tools seems to be inconclusive. Some interviewees clearly found cost of tools to be a prohibitive factor.}}} \\

  &       &       &       & \multicolumn{1}{p{10 cm}|}{{\scriptsize {\em  We have some problems with the complexity of the code generated. . . ``we are permanently optimizing this tool".}}} \\

  &       &       &       & \multicolumn{1}{p{10 cm}|}{{\scriptsize {\em  There is a danger, though, in believing that one ``killer application" of an MDE tool leads to another: ``prior to that they had used the technology successfully in a different project and it worked and they were very happy, so they thought, ok, this could be applied to virtually any kind of application".}}} \\

\hline

    \end{tabular}%
  \caption{Industrial issues evidenced (35/38).}
  \label{Table35}%
\end{table}%


\clearpage
\newpage

\begin{table}[htbp]
  \centering
    \begin{tabular}{|p{2 cm}|p{1 cm}|p{2 cm}|p{3 cm}|p{10 cm}|}
      \hline
    \multicolumn{1}{|c|}{{\bf Source}} & \multicolumn{1}{|c|}{{\bf  Year}}  & \multicolumn{1}{|c|}{{\bf Type}} & \multicolumn{1}{|c|}{{\bf Issue}} & \multicolumn{1}{|c|}{{\bf Statements}} \\
    \hline

    \multirow{15}[6]{*}{{\cite{whittle2013industrial}}} & \multirow{15}[6]{*}{2013} & \multirow{15}[6]{*}{\parbox{2cm}{Conference Paper}} & \multirow{5}[1]{*}{\parbox{3cm}{Tools as a way to increase complexity}} &  \multicolumn{1}{p{10 cm}|}{{\scriptsize {\em Vendors often spend a lot of time with clients customizing tools to a particular environment. But this can often cause delays and cost overruns and takes control away from the client: `` And suddenly the tool doesn’t do something expected and it’s a nightmare for them. So they try to contact the vendor but they do not really know what’s going on, they are mostly sales guys".}}} \\

  &       &       &       & \multicolumn{1}{p{10 cm}|}{{\scriptsize {\em  We recommend that the MDE community pay more attention to tried-and-tested HCI methods, which can help to produce more useful and usable tools.}}} \\

  &       &       &       & \multicolumn{1}{p{10 cm}|}{{\scriptsize {\em  MDE tools can make certification more difficult as current government certification processes are not set up to deal with auto-generated code.}}} \\

\cline{4-5}

  &       &       &    \multirow{5}[1]{*}{\parbox{3cm}{Implicit questions derived from the MDE adoption itself}}   & \\

  &       &       &     \multirow{9}[1]{*}{\parbox{3cm}{\&}}  & \multicolumn{1}{p{10 cm}|}{{\scriptsize {\em  An MDE effort started small, and was well supported by tools, but that processes and tools broke down when trying to roll out MDE across a wider part of the organization: ``the complexity of these little [DSL] languages started to grow and grow and grow. . . we were trying to share the [code generation] templates across teams and versioning and releasing of these templates was not under any kind of control at all"..}}} \\

  &       &       &    \multirow{1}[1]{*}{\parbox{3cm}{Tools as a way to increase complexity}}   & \\

  &       &       &      & \\
  &       &       &      & \\
\hline

    \end{tabular}%
  \caption{Industrial issues evidenced (36/38).}
  \label{Table36}%
\end{table}%


\clearpage
\newpage

\begin{table}[htbp]
  \centering
    \begin{tabular}{|p{2 cm}|p{1 cm}|p{2 cm}|p{3 cm}|p{10 cm}|}
      \hline
    \multicolumn{1}{|c|}{{\bf Source}} & \multicolumn{1}{|c|}{{\bf  Year}}  & \multicolumn{1}{|c|}{{\bf Type}} & \multicolumn{1}{|c|}{{\bf Issue}} & \multicolumn{1}{|c|}{{\bf Statements}} \\
    \hline

    \multirow{7}[6]{*}{{\cite{mora2006definicion}}} & \multirow{7}[6]{*}{2006} & \multirow{7}[6]{*}{\parbox{2cm}{Technical Report (in Spanish)}} & \multirow{8}[1]{*}{\parbox{3cm}{MDA is not enough}} &  \multicolumn{1}{p{10 cm}|}{{\scriptsize {\em MDA presenta carencias en transformaci\'on de modelos puesto que QVT a\'un est\'a en fase de aprobaci\'on como est\'andar OMG y no tiene soporte de herramientas (MDA has deficiencies in model transformations since QVT is still under approval as an OMG standard and it has not supporting tools.).}}} \\

  &       &       &       & \multicolumn{1}{p{10 cm}|}{{\scriptsize {\em MDA no proporciona suficientes gu\'{i}as metodol\'ogicas, s\'olo se define la estrategia general para la transformaci\'on PIM2PSM (MDA does not provide sufficient methodological guidelines, only it is defined an overall strategy for PIM-2-PSM transformation).}}} \\
  &       &       &       & \multicolumn{1}{p{10 cm}|}{{\scriptsize {\em MDA no garantiza la separaci\'on de conceptos (MDA does not guarantee the separation of concepts).}}} \\

\hline

    \multirow{10}[6]{*}{{\cite{Ortiz:2013:Online}}} & \multirow{10}[6]{*}{2013} & \multirow{10}[6]{*}{\parbox{2cm}{Web Page (in Spanish)}} & \multirow{3}[1]{*}{\parbox{3cm}{MDA is not enough}} &  \multicolumn{1}{p{10 cm}|}{{\scriptsize {\em MDA no implica la automatizaci\'on total de todos los procesos: podemos hacer transformaciones manualmente (MDA does not imply the complete automation of all processes: we can make changes manually).}}} \\

\cline{4-5}

  &       &       &      \multirow{5}[1]{*}{\parbox{3cm}{Tools as a way to increase complexity}}  & \multicolumn{1}{p{10 cm}|}{{\scriptsize {\em Todav\'{i}a poco soporte por herramientas (Still little support for tools).}}} \\

  &       &       &      & \\
  &       &       &      & \\
  &       &       &      & \\

\cline{4-5}

  &       &       &    \multirow{5}[1]{*}{\parbox{3cm}{MDA is not enough}}   & \\

  &       &       &     \multirow{6}[1]{*}{\parbox{3cm}{\&}}  & \multicolumn{1}{p{10 cm}|}{{\scriptsize {\em  Muchas de las herramientas no utilizan est\'andares MDA plenamente (ej.: UML2, QVT,etc.) (Many of the tools do not use fully MDA standards (eg., UML2, QVT, etc.)).}}} \\

  &       &       &    \multirow{5}[1]{*}{\parbox{3cm}{Tools as a way to increase complexity}}   & \\

  &       &       &      & \\
  &       &       &      & \\
  &       &       &      & \\
\hline

    \end{tabular}%
  \caption{Industrial issues evidenced (37/38).}
  \label{Table37}%
\end{table}%


\clearpage
\newpage

\begin{table}[htbp]
  \centering
    \begin{tabular}{|p{2 cm}|p{1 cm}|p{2 cm}|p{3 cm}|p{10 cm}|}
      \hline
    \multicolumn{1}{|c|}{{\bf Source}} & \multicolumn{1}{|c|}{{\bf  Year}}  & \multicolumn{1}{|c|}{{\bf Type}} & \multicolumn{1}{|c|}{{\bf Issue}} & \multicolumn{1}{|c|}{{\bf Statements}} \\
    \hline

    \multirow{15}[6]{*}{{\cite{igarza2012introduccion}}} & \multirow{15}[6]{*}{2012} & \multirow{15}[6]{*}{\parbox{2cm}{Journal Paper (in Spanish)}} & \multirow{5}[1]{*}{\parbox{3cm}{Tools as a way to increase complexity}} &  \multicolumn{1}{p{10 cm}|}{{\scriptsize {\em Pobre usabilidad de las herramientas (Poor usability of the tools).}}} \\

  &       &       &       & \multicolumn{1}{p{10 cm}|}{{\scriptsize {\em Problemas de interoperabilidad entre herramientas (Interoperability issues between tools).}}} \\

  &       &       &      & \\
  &       &       &      & \\
\cline{4-5}

   &  &  & \multirow{5}[1]{*}{\parbox{3cm}{Organizational support for the MDE adoption}} &  \multicolumn{1}{p{10 cm}|}{{\scriptsize {\em Falta concienciaci\'on sobre su potencial (Lack of awareness about its potential).}}} \\

   &  &  &  &  \multicolumn{1}{p{10 cm}|}{{\scriptsize {\em Dificultad de introducir nuevos m\'etodos en la empresa (Difficulty for introducing new methods in a company).}}} \\

   &  &  &  &  \multicolumn{1}{p{10 cm}|}{{\scriptsize {\em Mentalidad conservadora de muchos profesionales, inercia al cambio (Conservative beliefs of many professionals, inertia to change).}}} \\

   &  &  &   &  \multicolumn{1}{p{10 cm}|}{{\scriptsize {\em Empresas centradas en el corto plazo para conseguir el ROI (Companies focuse to achieve ROI in the short term).}}} \\

  &       &       &       & \multicolumn{1}{p{10 cm}|}{{\scriptsize {\em Dificultad de encontrar personas con experiencia (Difficulty to find people with expertice).}}} \\

\cline{4-5}

  &       &       &    \multirow{5}[1]{*}{\parbox{3cm}{MDA is not enough}}   & \\

  &       &       &     \multirow{6}[1]{*}{\parbox{3cm}{\&}}  & \multicolumn{1}{p{10 cm}|}{{\scriptsize {\em  Problemas de escalabilidad de los modelos (Scalability issues on the models).}}} \\

  &       &       &      & \multicolumn{1}{p{10 cm}|}{{\scriptsize {\em  Falta base te\'orica, tecnolog\'{i}a actual desarrollada ad-hoc (Lack of theoretical basis, current technology was developed in an ad-hoc way).}}} \\

  &       &       &    \multirow{5}[1]{*}{\parbox{3cm}{Implicit questions derived from the MDE adoption itself}}   & \\

  &       &       &      & \\
  &       &       &      & \\
  &       &       &      & \\
  &       &       &      & \\
\hline

    \end{tabular}%
  \caption{Industrial issues evidenced (38/38).}
  \label{Table38}%
\end{table}%

\end{landscape}

\chapter{Issues of research in modelling language quality evaluation}

This section presents some issues about quality in models and modelling languages traditionally addressed by academic/scientific communities.

\newpage

\begin{landscape}


\begin{table}[htbp]
  \centering
    \begin{tabular}{|p{2 cm}|p{1 cm}|p{2 cm}|p{3 cm}|p{9 cm}|}
      \hline
  	  \multicolumn{1}{|c|}{{\bf Source}} & \multicolumn{1}{|c|}{{\bf  Year}}  & \multicolumn{1}{|c|}{{\bf Type}} & \multicolumn{1}{|c|}{{\bf Issue}} & \multicolumn{1}{|c|}{{\bf Statements}} \\
    	\hline

	       \multirow{10}[1]{*}{\cite{Agner2013997}} & \multirow{10}[1]{*}{2013} & \multirow{10}[1]{*}{\parbox{2cm}{Journal paper}} & \multirow{5}[1]{*}{\parbox{2.8cm}{Defects and metrics mainly over UML}}  & \multicolumn{1}{p{10 cm}|}{{\scriptsize {\em The major problems encountered in the adoption of UML refer to the lack of skills, the lack of coherent tools, and the strict time 			requisites applicable to software development projects.}}} \\

 	&       &       &       &\\
	&       &       &       &\\
	&       &       &       &\\

 	 \cline{4-5}

	          &       &       & \multirow{6}[1]{*}{\parbox{3cm}{{\bf {\em Implicit questions derived from the MDE adoption itself}}}} & \multicolumn{1}{p{10 cm}|}{{\scriptsize {\em Software engineers are still not mature enough with regard to the use of MDE approaches.}}} \\

 	&       &       &       &\\
	&       &       &       &\\
	&       &       &       &\\
	&       &       &       &\\

	\hline

	 \multirow{10}[1]{*}{\cite{Panach2015338}} & \multirow{10}[1]{*}{2010} & \multirow{10}[1]{*}{\parbox{2cm}{Journal paper}} & \multirow{5}[1]{*}{\parbox{2.8cm}{Software quali-ty principles extrapolated at modeling
	levels}} & \multicolumn{1}{p{10 cm}|}{{\scriptsize {\em Usability is an important feature of systems, therefore MDD methods should provide a mechanism to abstractly represent this characteristic.}}} \\

 	&       &       &       &\\
	&       &       &       &\\
	&       &       &       &\\

\cline{4-5}

          &       &       & \multirow{5}[1]{*}{\parbox{2.8cm}{Specificity in the scenarios for quality in models}} & \multicolumn{1}{p{10 cm}|}{{\scriptsize {\em A shortcoming of these patterns is that each author deﬁnes the patterns with a different notation and a different syntax.}}} \\
 	&       &       &       &\\
	&       &       &       &\\
	&       &       &       &\\
  	  \hline

    \end{tabular}%
  \caption{Academic/scientific issues evidenced (1/14).}
  \label{Table19}%
\end{table}%


\newpage

\begin{table}[htbp]
  \centering
    \begin{tabular}{|p{2 cm}|p{1 cm}|p{2 cm}|p{3 cm}|p{9 cm}|}
      \hline
  	  \multicolumn{1}{|c|}{{\bf Source}} & \multicolumn{1}{|c|}{{\bf  Year}}  & \multicolumn{1}{|c|}{{\bf Type}} & \multicolumn{1}{|c|}{{\bf Issue}} & \multicolumn{1}{|c|}{{\bf Statements}} \\

    	\hline

     \multirow{10}[1]{*}{\cite{Fabra2012607}} & \multirow{10}[1]{*}{2012} & \multirow{10}[1]{*}{\parbox{2cm}{Journal paper}} & \multirow{5}[1]{*}{\parbox{2.8cm}{Specificity in the scenarios for quality in models}} & \multicolumn{1}{p{10 cm}|}{{\scriptsize {\em It is difﬁcult for business analysts to use them in the early stages of the development process (Verner, 2004) (analysis and modelling stages), their dependence speciﬁc implementation technology (more speciﬁcally, on the service technology) and, ﬁnally, the lack of formal semantics that permit process analysis. These problems increase the gap between business analysts and software developers, which represents a serious limitation in the BPM field. on a Web
 }}} \\
    \cline{4-5}
          &       &       & \multirow{5}[1]{*}{\parbox{2.8cm}{Hard operationalization of model-quality frameworks}} & \multicolumn{1}{p{10 cm}|}{{\scriptsize {\em However, the existence of multiple schemas and separate graphics complicates validation, and the graphic portability in existing tools is becoming more difﬁcult.}}} \\

	&       &       &       &\\
	&       &       &       &\\

\hline

    \multirow{8}[1]{*}{\cite{5298434}} & \multirow{8}[1]{*}{2009} & \multirow{8}[1]{*}{\parbox{2cm}{Conference Paper}} & \multirow{5}[1]{*}{\parbox{2.8cm}{{\bf {\em Tools as a way to increase complexity}}}} & \multicolumn{1}{p{10 cm}|}{{\scriptsize {\em It was also noticed that some of the problems were caused by the use of immature tools, causing, for
example, the explosion of models if the metamodel is changed.}}} \\

	&       &       &       &\\
	&       &       &       &\\

    \cline{4-5}
          &       &       & 	 \multirow{5}[1]{*}{\parbox{2.8cm}{Hard operationalization of model-quality frameworks}} & \multicolumn{1}{p{10 cm}|}{{\scriptsize {\em Many of these interrelationships are complex. The round-trip problem occurs whenever an interrelated artifact changes in ways that affect some or all of its related artifacts as the mutual consistency cannot always be automatically assured.}}}  \\
	&       &       &       &\\
    \hline

    \end{tabular}%
  \caption{Academic/scientific issues evidenced (2/14).}
  \label{Table20}%
\end{table}%


\newpage

\begin{table}[htbp]
  \centering
    \begin{tabular}{|p{2 cm}|p{1 cm}|p{2 cm}|p{3 cm}|p{9 cm}|}
      \hline
  	  \multicolumn{1}{|c|}{{\bf Source}} & \multicolumn{1}{|c|}{{\bf  Year}}  & \multicolumn{1}{|c|}{{\bf Type}} & \multicolumn{1}{|c|}{{\bf Issue}} & \multicolumn{1}{|c|}{{\bf Statements}} \\

    	\hline

      \multirow{5}[1]{*}{\cite{Davies201588}} & \multirow{5}[1]{*}{2014}  & \multirow{5}[1]{*}{\parbox{2cm}{Journal paper}} & \multirow{5}[1]{*}{\parbox{2.8cm}{Defects and metrics mainly over UML}}  & \multicolumn{1}{p{10 cm}|}{{\scriptsize {\em We may observe that the behavioural semantics for UML is not compositional with regard to abstraction and concurrency: that is, if we allow concurrent execution of operations, then it is not possible to derive a behavioural speciﬁcation of a compound model from the behavioural speciﬁcations of its components—unless that speciﬁcation contains every detail  of the implementation, and thus has no abstraction at all.}}} \\
    \hline

  \multirow{5}[1]{*}{\cite{Panach2015164}} & \multirow{5}[1]{*}{2015}  & \multirow{5}[1]{*}{\parbox{2cm}{Journal paper}}  & \multirow{7}[1]{*}{\parbox{2.8cm}{{\bf {\em Implicit questions derived from the MDE adoption itself}}}} & \multicolumn{1}{p{10 cm}|}{{\scriptsize {\em For small systems and less programming-experienced subjects, MDD does not always yield better results than a traditional method, even regarding effort and productivity. This contradicts some previous statements about MDD advantages. The beneﬁts of developing a system with MDD appear to depend on certain characteristics of the development context.}}} \\
	
	&       &       &       &\\
	&       &       &       &\\

\hline

 \multirow{9}[1]{*}{\cite{Molina20091306}}  & \multirow{9}[1]{*}{2009} & \multirow{9}[1]{*}{\parbox{2cm}{Journal paper}} & \multirow{6}[1]{*}{\parbox{2.8cm}{Hard operationalization of model-quality frameworks}} &  \multicolumn{1}{p{10 cm}|}{{\scriptsize {\em These studies highlight that the top five problem areas of large-scale WE projects are (1) failure
to meet business needs (84\%), (2) project schedule delays (79\%), (3) budget overrun (63\%), (4) lack of required functionality (53\%) and (5) poor quality of deliverables (52\%).}}} \\

&       &       &       &\\
&       &       &       &\\

\cline{4-5}
          &       &       & \multirow{5}[1]{*}{\parbox{2.8cm}{Software quali-ty principles extrapolated at modeling levels}} & \multicolumn{1}{p{10 cm}|}{{\scriptsize {\em The main problem of the ﬁrst option is that it creates a hard coupling between the elements of which the navigational metamodels were originally composed and the new added elements. This might limit the extensibility of the proposal and it would be difﬁcult for modellers to distinguish between the elements of the original metamodel and the elements added to support usability features. }}} \\
    \hline

    \end{tabular}%
  \caption{Academic/scientific issues evidenced (3/14).}
  \label{Table21}%
\end{table}%


\newpage

\begin{table}[htbp]
  \centering
    \begin{tabular}{|p{2 cm}|p{1 cm}|p{2 cm}|p{3 cm}|p{9 cm}|}
      \hline
  	  \multicolumn{1}{|c|}{{\bf Source}} & \multicolumn{1}{|c|}{{\bf  Year}}  & \multicolumn{1}{|c|}{{\bf Type}} & \multicolumn{1}{|c|}{{\bf Issue}} & \multicolumn{1}{|c|}{{\bf Statements}} \\

    	\hline

    \multirow{20}[1]{*}{\cite{Nugroho20091670}} & \multirow{20}[1]{*}{2009} & \multirow{20}[1]{*}{\parbox{2cm}{Journal paper}} & \multirow{5}[1]{*}{\parbox{2.8cm}{Defects and metrics mainly over UML}}  & \multicolumn{1}{p{10 cm}|}{{\scriptsize {\em In this line of research, the focus has been on the formality of UML models and its relation with model quality and comprehensibility. A previous study that deserves attention is the one from Briand et al. In their experimental study, Briand et al. investigated the impact of using OCL (object constraint language) in UML models on defect detection, comprehension, and impact analysis of changes.}}} \\

          &       &       & & \multicolumn{1}{p{10 cm}|}{{\scriptsize {\em UML models with low LoD generally have lower comprehensibility because they are often misunderstood or misinterpreted by the readers.}}} \\
          &       &       &       & \multicolumn{1}{p{10 cm}|}{{\scriptsize {\em Therefore, software designers should be aware of the trade-off and subsequently make informed decisions to target the quality levels of their models.}}} \\
          &       &       &       & \multicolumn{1}{p{10 cm}|}{{\scriptsize {\em A study from Arisholm et al. looked at the problem from a coarser grained view: the absence/presence of UML in software maintenance. The results conﬁrmed that the use of UML for maintenance signiﬁcantly reduces time to make code changes in the system and increases functional correctness of the changes. However, the authors also stated that effort saving was not visible when the time required to change the UML diagrams was taken into consideration.}}} \\
\cline{4-5}
          &       &       & \multirow{5}[1]{*}{\parbox{2.8cm}{Hard operationalization of model-quality frameworks}} & \multicolumn{1}{p{10 cm}|}{{\scriptsize {\em informal modeling styles can safe time and effort, but might lead to problems related to interpretations of the models.}}} \\
&       &       &       &\\
&       &       &       &\\
&       &       &       &\\
\hline

 \multirow{3}[1]{*}{\cite{Gorschek2014176}} & \multirow{3}[1]{*}{2014}  & \multirow{3}[1]{*}{\parbox{2cm}{Journal paper}} & \multirow{4}[1]{*}{\parbox{2.8cm}{Defects and metrics mainly over UML}}  & \multicolumn{1}{p{10 cm}|}{{\scriptsize {\em I use UML to visualize and design the class structure, but once  coding starts to gain momentum, the UML is left behind}}} \\
&       &       &       &\\
&       &       &       &\\
    \hline

    \end{tabular}%
  \caption{Academic/scientific issues evidenced (4/14).}
  \label{Table22}%
\end{table}%


\newpage

\begin{table}[htbp]
  \centering
    \begin{tabular}{|p{2 cm}|p{1 cm}|p{2 cm}|p{3 cm}|p{9 cm}|}
      \hline
  	  \multicolumn{1}{|c|}{{\bf Source}} & \multicolumn{1}{|c|}{{\bf  Year}}  & \multicolumn{1}{|c|}{{\bf Type}} & \multicolumn{1}{|c|}{{\bf Issue}} & \multicolumn{1}{|c|}{{\bf Statements}} \\

    	\hline

     \multirow{5}[1]{*}{\cite{6604438}} & \multirow{5}[1]{*}{2013} & \multirow{5}[1]{*}{\parbox{2cm}{Workshop Paper}} & \multirow{5}[1]{*}{\parbox{2.8cm}{Hard operationalization of model-quality frameworks}} & \multicolumn{1}{p{10 cm}|}{{\scriptsize {\em Although, Search-based Software Engineering (SBSE)  has been successfully applied to a number of different MDE tasks, such as model transformation, model evolution, model analysis, and model transformation testing, applying SBSE to complex MDE problems necessitates expertise in both, search-based optimization algorithms and MDE formalisms and techniques. }}} \\
    \hline
 \multirow{3}[1]{*}{\cite{LagunaMarques}} & \multirow{3}[1]{*}{2010}  & \multirow{3}[1]{*}{\parbox{2cm}{Journal paper}} & \multirow{3}[1]{*}{\parbox{2.8cm}{ Defects and metrics mainly over UML}}  & \multicolumn{1}{p{10 cm}|}{{\scriptsize {\em On the other hand, the variants in UML models can be distributed between several modeling elements. The problem is due to the fact that a variable feature can correspond to several elements in a UML model }}} \\

&       &       &       &\\

\hline

 \multirow{5}[1]{*}{\cite{Wehrmeister2014844}} & \multirow{5}[1]{*}{2014} & \multirow{5}[1]{*}{\parbox{2cm}{Journal paper}} & \multirow{5}[1]{*}{\parbox{2.8cm}{Defects and metrics mainly over UML }} & \multicolumn{1}{p{10 cm}|}{{\scriptsize {\em According to the authors, Theme/UML has some limitations: (i) it addresses only non-functional concerns that manifest themselves as code in the system; (ii) inheritance is not supported in UML-to-C transformation; and (iii) the behavior speciﬁed within aspects can be speciﬁed only with sequence diagram, leading to composition problems when state diagrams are used to specify functional requirements.}}} \\

    \hline

    \end{tabular}%
  \caption{Academic/scientific issues evidenced (5/14).}
  \label{Table23}%
\end{table}%


\clearpage
\newpage

\begin{table}[htbp]
  \centering
    \begin{tabular}{|p{2 cm}|p{1 cm}|p{2 cm}|p{3 cm}|p{9 cm}|}
      \hline
  	  \multicolumn{1}{|c|}{{\bf Source}} & \multicolumn{1}{|c|}{{\bf  Year}}  & \multicolumn{1}{|c|}{{\bf Type}} & \multicolumn{1}{|c|}{{\bf Issue}} & \multicolumn{1}{|c|}{{\bf Statements}} \\

    	\hline

  \multirow{10}[1]{*}{\cite{Dijkman20081281}} & \multirow{10}[1]{*}{2009} & \multirow{10}[1]{*}{\parbox{2cm}{Journal paper}} & \multirow{5}[1]{*}{\parbox{2.8cm}{Hard operationalization of model-quality frameworks}} & \multicolumn{1}{p{10 cm}|}{{\scriptsize {\em It is known that models deﬁned in this family of languages may exhibit a range of semantic errors, including deadlocks and livelocks. Such errors are especially problematic at the levels of domain analysis and high-level systems design, because errors at these levels are among the hardest and most costly to correct.}}}  \\

\cline{4-5}
          &       &       & \multirow{5}[1]{*}{\parbox{2.8cm}{Defects and metrics mainly over UML}} & \multicolumn{1}{p{10 cm}|}{{\scriptsize {\em The lack of formal semantics of BPMN hinders on the development of tool support for checking the correctness of BPMN models from a semantic perspective.}}} \\
          &       &       &       & \multicolumn{1}{p{10 cm}|}{{\scriptsize {\em The BPMN standard speciﬁcation is relatively detailed when it comes to specifying syntactic constraints on BPMN models, but it is unsystematic and sometimes inconsistent when it comes to deﬁning their semantics.}}}  \\
    \hline

  \multirow{7}[1]{*}{\cite{Davies2006358}} & \multirow{7}[1]{*}{2004} & \multirow{7}[1]{*}{\parbox{2cm}{Journal paper}} & \multirow{7}[1]{*}{\parbox{2.8cm}{{\bf {\em  Implicit questions derived from the MDE adoption itself}}}} & \multicolumn{1}{p{10 cm}|}{{\scriptsize {\em Clearly, relative advantage (disadvantage)/usefulness from the perspective of the analyst was the major driving factor influencing the decision to continue (discontinue) modeling.... this study identified five factors that uniquely influence the continued use decision of analysts, viz., communication (using diagrams) to/from stakeholders, internal knowledge (lack of) of techniques, user expectations management, understanding models integration into the business, and tool/software deficiencies}}}  \\

\hline

    \end{tabular}%
  \caption{Academic/scientific issues evidenced (6/14).}
  \label{Table24}%
\end{table}%


\clearpage
\newpage

\begin{table}[htbp]
  \centering
    \begin{tabular}{|p{2 cm}|p{1 cm}|p{2 cm}|p{3 cm}|p{9 cm}|}
      \hline
  	  \multicolumn{1}{|c|}{{\bf Source}} & \multicolumn{1}{|c|}{{\bf  Year}}  & \multicolumn{1}{|c|}{{\bf Type}} & \multicolumn{1}{|c|}{{\bf Issue}} & \multicolumn{1}{|c|}{{\bf Statements}} \\

    	\hline

  \multirow{18}[1]{*}{\cite{mussbacher2014relevance}} & \multirow{18}[1]{*}{2014} & \multirow{18}[1]{*}{\parbox{2cm}{Conference Paper}} &\multirow{18}[1]{*}{\parbox{2.8cm}{{\bf {\em  Implicit questions derived from the MDE adoption itself}}}} & \multicolumn{1}{p{10 cm}|}{{\scriptsize {\em {\bf (Shortcomings of MDE to Address Increasing Demands on Software)}. Often, mature tools provide techniques that can successfully cope with software systems that we were building a decade ago, but fail when applied to model complex systems like the ones described above.
}}}  \\

          &       &       &  & \multicolumn{1}{p{10 cm}|}{{\scriptsize {\em {\bf (Models Are Still Not Valued as Much as Code)}. Unfortunately, for many people, modeling is considered a superfluous activity that becomes an activity in itself not necessarily for the benefit of the software development.
}}} \\
          &       &       &  & \multicolumn{1}{p{10 cm}|}{{\scriptsize {\em {\bf (Lack of Fundamentals in MDE)}. Unlike most other fields of engineering, model driven engineering does not have a Body of Knowledge (BoK) as such.}}} \\

          &       &       &  & \multicolumn{1}{p{10 cm}|}{{\scriptsize {\em {\bf (Education Issues)}. For students, it is difficult to learn to use their abstraction abilities ..., which have been shown to closely relate to software design skills.}}} \\

          &       &       &  & \multicolumn{1}{p{10 cm}|}{{\scriptsize {\em {\bf (Lack of (Industrial) Evidence of Benefits)}. We are still lacking knowledge on factors that make MDE successful, also considering that model-based approaches are regularly used in the hardware industry (e.g., model checking to analyze hardware designs instead of testing).}}} \\

          &       &       &  & \multicolumn{1}{p{10 cm}|}{{\scriptsize {\em {\bf (MDE Is Not Considered ``Cool")}.Even though MDE has been around for over 10 years, it is currently not as widespread in industry as the modeling community has hoped.}}} \\

\hline

    \end{tabular}%
  \caption{Academic/scientific issues evidenced (7/14).}
  \label{Table24}%
\end{table}%


\clearpage
\newpage

\begin{table}[htbp]
  \centering
    \begin{tabular}{|p{2 cm}|p{1 cm}|p{2 cm}|p{3 cm}|p{9 cm}|}
      \hline
  	  \multicolumn{1}{|c|}{{\bf Source}} & \multicolumn{1}{|c|}{{\bf  Year}}  & \multicolumn{1}{|c|}{{\bf Type}} & \multicolumn{1}{|c|}{{\bf Issue}} & \multicolumn{1}{|c|}{{\bf Statements}} \\

    	\hline

  \multirow{18}[1]{*}{\cite{mussbacher2014relevance}} & \multirow{18}[1]{*}{2014} & \multirow{18}[1]{*}{\parbox{2cm}{Conference Paper}} &\multirow{7}[1]{*}{\parbox{2.8cm}{{\bf {\em Tools as a way to increase complexity}}}} & \multicolumn{1}{p{10 cm}|}{{\scriptsize {\em {\bf (Obstacles for Tool Usability and Adoption)}. The users face significant usability challenges ..., e.g., steep learning curves, arduous user interfaces, and difficulty with migrating models from one version of a tool to the next.}}}  \\

          &       &       &  & \multicolumn{1}{p{10 cm}|}{{\scriptsize {\em Tools do not support the fundamentally creative side of the modeling process due to their inflexibility and complexity. }}} \\

          &       &       &  & \multicolumn{1}{p{10 cm}|}{{\scriptsize {\em Finally, model transformations, which are essential in order for MDE to be effective, are difficult to maintain and adapt to changing requirements and implementation platforms.}}} \\

\cline{4-5}

  &       &       &    \multirow{5}[1]{*}{\parbox{3cm}{{\bf {\em MDA is not enough}}}}   & \\

  &       &       &    & \\
  &       &       &    & \\

  &       &       &     \multirow{6}[1]{*}{\parbox{3cm}{\&}}  & \multicolumn{1}{p{10 cm}|}{{\scriptsize {\em  {\bf (Inconsistencies between Software Artifacts)}. A complicating factor is that often a system is modeled with multiple views using different models and modeling notations, thus further increasing the likelihood of introducing inconsistencies between these models. Even when additional information is overlaid onto an existing view (as is the case, for example, in UML, when stereotypes define non-functional properties), there are no guarantees that the resulting system is consistent or correctly functioning.}}} \\

  &       &       &    \multirow{1}[1]{*}{\parbox{3cm}{{\bf {\em Implicit questions derived from the MDE adoption itself}}}}   & \\

  &       &       &    & \\
  &       &       &    & \\
  &       &       &    & \\
  &       &       &    & \\
\hline

    \end{tabular}%
  \caption{Academic/scientific issues evidenced (8/14).}
  \label{Table24}%
\end{table}%


\clearpage
\newpage

\begin{table}[htbp]
  \centering
    \begin{tabular}{|p{2 cm}|p{1 cm}|p{2 cm}|p{3 cm}|p{9 cm}|}
      \hline
  	  \multicolumn{1}{|c|}{{\bf Source}} & \multicolumn{1}{|c|}{{\bf  Year}}  & \multicolumn{1}{|c|}{{\bf Type}} & \multicolumn{1}{|c|}{{\bf Issue}} & \multicolumn{1}{|c|}{{\bf Statements}} \\

    	\hline

  \multirow{9}[1]{*}{\cite{poruban2014pragmatic}} & \multirow{9}[1]{*}{2014} & \multirow{9}[1]{*}{\parbox{2cm}{Conference Paper}} &\multirow{7}[1]{*}{\parbox{2.8cm}{{\bf {\em Tools as a way to increase complexity}}}} & \multicolumn{1}{p{10 cm}|}{{\scriptsize {\em }}}  \\

  &       &       &    & \\
  &       &       &    & \\

  &       &       &     \multirow{6}[1]{*}{\parbox{3cm}{\&}}  & \multicolumn{1}{p{10 cm}|}{{\scriptsize {\em  MDSD requires massive tool support. These tools, however, are often complex and require high learning costs}}} \\

  &       &       &    & \\
  &       &       &    & \\

  &       &       &    \multirow{1}[1]{*}{\parbox{3cm}{{\bf {\em Implicit questions derived from the MDE adoption itself}}}}   & \\

  &       &       &    & \\
  &       &       &    & \\
  &       &       &    & \\
  &       &       &    & \\
\hline

  \multirow{9}[1]{*}{\cite{bruel2015mde}} & \multirow{9}[1]{*}{2015} & \multirow{9}[1]{*}{\parbox{2cm}{Journal Paper}} &\multirow{7}[1]{*}{\parbox{2.8cm}{{\bf {\em Tools as a way to increase complexity}}}} & \multicolumn{1}{p{10 cm}|}{{\scriptsize {\em The need for mature, design-oriented, user-friendly tools is naturally not specific to the scientific computing and it is often cited as a hinder in a faster (wider) adoption of these techniques in the industry.}}}  \\

  &       &       &    &  \multicolumn{1}{p{10 cm}|}{{\scriptsize {\em Not clear what such a tool support actually means, but it is clear that it should at least include an easy install procedure, integrated versioning mechanisms and support for collaborative work (such as web based platforms).}}}  \\

\cline{4-5}

   &  &  &\multirow{7}[1]{*}{\parbox{2.8cm}{{\bf {\em Implicit questions derived from the MDE adoption itself}}}} & \multicolumn{1}{p{10 cm}|}{{\scriptsize {\em Another challenge for the MDE community is to support design as an art, not just as a high-level programming technique. More efforts, both on methodology (for collaboratively building conceptual models) and on tools (more intuitive and less computer-oriented) would certainly help.}}}  \\

  &       &       &    & \\
  &       &       &    & \\
  &       &       &    & \\

\hline

    \end{tabular}%
  \caption{Academic/scientific issues evidenced (9/14).}
  \label{Table24}%
\end{table}%


\clearpage
\newpage

\begin{table}[htbp]
  \centering
    \begin{tabular}{|p{2 cm}|p{1 cm}|p{2 cm}|p{3 cm}|p{9 cm}|}
      \hline
  	  \multicolumn{1}{|c|}{{\bf Source}} & \multicolumn{1}{|c|}{{\bf  Year}}  & \multicolumn{1}{|c|}{{\bf Type}} & \multicolumn{1}{|c|}{{\bf Issue}} & \multicolumn{1}{|c|}{{\bf Statements}} \\

    	\hline

  \multirow{12}[1]{*}{\cite{picek2007model}} & \multirow{12}[1]{*}{2007} & \multirow{12}[1]{*}{\parbox{2cm}{Conference Paper}} &\multirow{7}[1]{*}{\parbox{2.8cm}{{\bf {\em Implicit questions derived from the MDE adoption itself}}}} & \multicolumn{1}{p{10 cm}|}{{\scriptsize {\em Some authors ... believe that MDD has a chance to succeed in today's software industry, but still it is far from a sure bet,}}}  \\

  &       &       &    & \multicolumn{1}{p{10 cm}|}{{\scriptsize {\em  }}}\\
  &       &       &    & \multicolumn{1}{p{10 cm}|}{{\scriptsize {\em  }}}\\
  &       &       &    & \multicolumn{1}{p{10 cm}|}{{\scriptsize {\em  }}}\\
  &       &       &    & \multicolumn{1}{p{10 cm}|}{{\scriptsize {\em  }}}\\
  &       &       &    & \multicolumn{1}{p{10 cm}|}{{\scriptsize {\em  }}}\\

\cline{4-5}

  &       &       &  \multirow{7}[1]{*}{\parbox{3cm}{Defects and metrics mainly over UML}} & \multicolumn{1}{p{10 cm}|}{{\scriptsize {\em   To be able to address so many needs, UML 2.0 becomes enormous, ambiguous and unwieldy. It contains some diagrams and constructs that are redundant or infrequently used.}}}\\

  &       &       &    & \multicolumn{1}{p{10 cm}|}{{\scriptsize {\em  UML 2.0 lacks a reference implementation and a human-readable semantic account to provide an operational semantics, so it's difficult to interpret and correctly implement UML model transformation tools.}}}\\

  &       &       &    & \multicolumn{1}{p{10 cm}|}{{\scriptsize {\em The lack of semantics precision makes the production of automated MDD tools difficult because the semantics carries the meaning that is essential to enable automation.}}}\\

  &       &       &    & \multicolumn{1}{p{10 cm}|}{{\scriptsize {\em  Clearly, UML or any other MDD language faces significant hurdles to demonstrate sufficient value to satisfy the needs of all the different kinds of MDD users.}}}\\

\hline

    \end{tabular}%
  \caption{Academic/scientific issues evidenced (10/14).}
  \label{Table24}%
\end{table}%


\clearpage
\newpage

\begin{table}[htbp]
  \centering
    \begin{tabular}{|p{2.2 cm}|p{1 cm}|p{2 cm}|p{3 cm}|p{9 cm}|}
      \hline
  	  \multicolumn{1}{|c|}{{\bf Source}} & \multicolumn{1}{|c|}{{\bf  Year}}  & \multicolumn{1}{|c|}{{\bf Type}} & \multicolumn{1}{|c|}{{\bf Issue}} & \multicolumn{1}{|c|}{{\bf Statements}} \\

    	\hline

  \multirow{20}[1]{*}{\small {\cite{van2008challenges}}} & \multirow{20}[1]{*}{2008} & \multirow{20}[1]{*}{\parbox{2cm}{Conference Paper}} &\multirow{20}[1]{*}{\parbox{2.8cm}{ Specificity in the scenarios for quality in models}} & \multicolumn{1}{p{10 cm}|}{{\scriptsize {\em MDE is still lacking adoption by developers. To live up to its full potential MDE must rest on a solid foundation. Therefore, one of the main challenges facing MDE today is the establishment of such a foundation.}}}  \\

  &       &       &    & \multicolumn{1}{p{10 cm}|}{{\scriptsize {\em  Firstly, the specification of the behavioural semantics of meta-models so that different kinds of analysis can be conducted, e.g., simulation, validation and model checking.}}}\\
  &       &       &    & \multicolumn{1}{p{10 cm}|}{{\scriptsize {\em   A second challenge is the support of the notion of time in these behavioural descriptions, to be able to conduct, e.g., realistic performance and reliability analysis of industrial systems.}}}\\
  &       &       &    & \multicolumn{1}{p{10 cm}|}{{\scriptsize {\em   As a third challenge, not only the accidental complexity involved in building software systems needs to be tackled, but their essential complexity should be addressed too.}}}\\
  &       &       &    & \multicolumn{1}{p{10 cm}|}{{\scriptsize {\em  Some promising approaches revolve around the satisfability property of a model, i.e., deciding whether it is possible to create a well-formed instantiation of the model. Existing solutions in the UML/OCL context were discussed. The presenter claimed that this problem has not yet been satisfactorily addressed.}}}\\
  &       &       &    & \multicolumn{1}{p{10 cm}|}{{\scriptsize {\em  The presenter argued for the necessity of extending existing methods with improved requirement techniques based on goal-oriented techniques for the analysis and specification of the organisation context, and discussed the benefits and challenges of such integration.}}}\\
  &       &       &    & \multicolumn{1}{p{10 cm}|}{{\scriptsize {\em  Concerning the intrinsic type of scalability needed for MDE, one of the main problems is that MDE has to be able to cope with very large models in order to model systems of systems and Ultra-Large-Scale (ULS) systems.}}}\\
  &       &       &    & \multicolumn{1}{p{10 cm}|}{{\scriptsize {\em  The question arises whether generic optimisation solutions can be developed for MDE activities. In addition, one must be aware that the time to load huge models is often greater than the time needed for checking, merging or transforming such models.}}}\\

\hline

    \end{tabular}%
  \caption{Academic/scientific issues evidenced (11/14).}
  \label{Table24}%
\end{table}%


\clearpage
\newpage

\begin{table}[htbp]
  \centering
    \begin{tabular}{|p{2.2 cm}|p{1 cm}|p{2 cm}|p{3.5 cm}|p{9 cm}|}
      \hline
  	  \multicolumn{1}{|c|}{{\bf Source}} & \multicolumn{1}{|c|}{{\bf  Year}}  & \multicolumn{1}{|c|}{{\bf Type}} & \multicolumn{1}{|c|}{{\bf Issue}} & \multicolumn{1}{|c|}{{\bf Statements}} \\

    	\hline

  \multirow{5}[1]{*}{{\cite{kolovos2008grand}}} & \multirow{5}[1]{*}{2008} & \multirow{5}[1]{*}{\parbox{2cm}{Conference Paper}} &\multirow{5}[1]{*}{\parbox{3.5cm}{Software quality principles extrapolated at modeling levels}} & \multicolumn{1}{p{10 cm}|}{{\scriptsize {\em MDE should go beyond conceptual modeling and generative programming: it should count on mature tool-support for automating the design, development and analysis of systems, as well as on measurable engineering processes and methodologies to drive the efective use of all these artifacts towards the predictable construction of software systems.}}}  \\

\hline

  \multirow{9}[1]{*}{\cite{rivera2008behavior}} & \multirow{9}[1]{*}{2008} & \multirow{9}[1]{*}{\parbox{2cm}{Conference Paper}} &\multirow{7}[1]{*}{\parbox{2.8cm}{Hard operationalization of model-quality frameworks}} & \multicolumn{1}{p{10 cm}|}{{\scriptsize {\em Firstly, the specification of the behavioral semantics of metamodels (beyond their basic structure), so that diferent kinds of analysis can be conducted, e.g., simulation, validation and model checking.}}}  \\

  &       &       &     \multirow{6}[1]{*}{\parbox{3cm}{\&}}  & \multicolumn{1}{p{10 cm}|}{{\scriptsize {\em  Second challenge is the support of the notion of time in these behavioral descriptions, another key issue to allow industrial systems to be realistically simulated and properly analyzed|to be able to conduct e.g., performance and reliability analysis.}}} \\

  &       &       &    \multirow{1}[1]{*}{\parbox{3cm}{Specificity in the scenarios for quality in models}}   &  \multicolumn{1}{p{10 cm}|}{{\scriptsize {\em  Finally, we need not only to tackle the accidental complexity involved building software systems, but we should also try to deal with their essential comple-xity.}}} \\

  &       &       &    & \\

\hline

    \end{tabular}%
  \caption{Academic/scientific issues evidenced (12/14).}
  \label{Table24}%
\end{table}%


\clearpage
\newpage

\begin{table}[htbp]
  \centering
    \begin{tabular}{|p{2 cm}|p{1 cm}|p{2 cm}|p{3 cm}|p{9 cm}|}
      \hline
  	  \multicolumn{1}{|c|}{{\bf Source}} & \multicolumn{1}{|c|}{{\bf  Year}}  & \multicolumn{1}{|c|}{{\bf Type}} & \multicolumn{1}{|c|}{{\bf Issue}} & \multicolumn{1}{|c|}{{\bf Statements}} \\

    	\hline

  \multirow{20}[1]{*}{\cite{quintero2011reflexiones}} & \multirow{20}[1]{*}{2011} & \multirow{20}[1]{*}{\parbox{2cm}{Journal Paper (in Spanish)}} &\multirow{12}[1]{*}{\parbox{2.8cm}{{\bf {\em Implicit questions derived from the MDE adoption itself}}}} & \multicolumn{1}{p{10 cm}|}{{\scriptsize {\em Los modelos se vuelven obsoletos e incompatibles con el c\'odigo. (Models become obsolete and incompatible with the code).}}}  \\

  &       &       &    & \multicolumn{1}{p{10 cm}|}{{\scriptsize {\em Los modelos no se pueden intercambiar f\'acilmente entre las herramientas. (Models cannot switch between tools easily).}}} \\

  &       &       &     \multirow{13}[1]{*}{\parbox{3cm}{\&}}  & \multicolumn{1}{p{10 cm}|}{{\scriptsize {\em Las herramientas de modelado son "peso pesado" al instalar, aprender, configurar y utilizar(Modelling tools are ``heavyweight" to install, learn, configure and use).}}} \\

  &       &       &    & \multicolumn{1}{p{10 cm}|}{{\scriptsize {\em El c\'odigo generado por una herramienta de modelado no es del tipo que me gustar\'{i}a (The generated code of a modelling tool is not the kind I would like).}}} \\

  &       &       &    \multirow{10}[1]{*}{\parbox{3cm}{{\bf {\em Tools as a way to increase complexity}}}}   & \multicolumn{1}{p{10 cm}|}{{\scriptsize {\em No se puede describir el tipo de detalles que deben ser implementados (The kind of details that must be implemented cannot be described).}}} \\

  &       &       &    & \multicolumn{1}{p{10 cm}|}{{\scriptsize {\em Algunas de las herramientas que inicialmente parec\'{i}an bastante prometedoras en el marco del MDA, como ArcStyler y OptimalJ (Garc\'{i}a et al., 2004), que incluso estaban incluidas en el listado oficial del Object Management Group, han perdido su auge, al punto de que en algunos casos las compa\~n\'{i}as encargadas tuvieron que cambiar de estrategia de negocio frente al tema (Some of the tools that initially seemed quite promising with the MDA framework, such as ArcStyler and OptimalJ, which were even included in the official list of the Object Management Group, lost their boom. In some cases the companies had to change their business strategy regarding this field).}}} \\

  &       &       &    & \\

  &       &       &    & \\
\hline

    \end{tabular}%
  \caption{Academic/scientific issues evidenced (13/14).}
  \label{Table24}%
\end{table}%


\clearpage
\newpage

\begin{table}[htbp]
  \centering
    \begin{tabular}{|p{2 cm}|p{1 cm}|p{2 cm}|p{3 cm}|p{9 cm}|}
      \hline
  	  \multicolumn{1}{|c|}{{\bf Source}} & \multicolumn{1}{|c|}{{\bf  Year}}  & \multicolumn{1}{|c|}{{\bf Type}} & \multicolumn{1}{|c|}{{\bf Issue}} & \multicolumn{1}{|c|}{{\bf Statements}} \\

    	\hline

  \multirow{12}[1]{*}{\cite{quintero2012face}} & \multirow{12}[1]{*}{2012} & \multirow{12}[1]{*}{\parbox{2cm}{Conference Paper}} &\multirow{7}[1]{*}{\parbox{2.8cm}{{\bf {\em Implicit questions derived from the MDE adoption itself}}}} & \multicolumn{1}{p{10 cm}|}{{\scriptsize {\em Models become out of date and inconsistent with code.}}}  \\

  &       &       &    & \multicolumn{1}{p{10 cm}|}{{\scriptsize {\em  }}}\\
  &       &       &    & \multicolumn{1}{p{10 cm}|}{{\scriptsize {\em  }}}\\
  &       &       &    & \multicolumn{1}{p{10 cm}|}{{\scriptsize {\em  }}}\\
  &       &       &    & \multicolumn{1}{p{10 cm}|}{{\scriptsize {\em  }}}\\
  &       &       &    & \multicolumn{1}{p{10 cm}|}{{\scriptsize {\em  }}}\\
  &       &       &    & \multicolumn{1}{p{10 cm}|}{{\scriptsize {\em  }}}\\

\cline{4-5}

  &       &       &  \multirow{7}[1]{*}{\parbox{2.8cm}{{\bf {\em Tools as a way to increase complexity}}}} & \multicolumn{1}{p{10 cm}|}{{\scriptsize {\em   Models cannot be easily exchanged between tools.}}}\\

  &       &       &    & \multicolumn{1}{p{10 cm}|}{{\scriptsize {\em  Modeling tools are ``heavyweight" (to install, learn, configure, use).}}}\\

  &       &       &    & \multicolumn{1}{p{10 cm}|}{{\scriptsize {\em Code generated from a modeling tool is not of the kind I would like.}}}\\

  &       &       &    & \multicolumn{1}{p{10 cm}|}{{\scriptsize {\em  Modeling tools change, models become obsolete.}}}\\

  &       &       &    & \multicolumn{1}{p{10 cm}|}{{\scriptsize {\em Modeling tools are too expensive.}}}\\

  &       &       &    & \multicolumn{1}{p{10 cm}|}{{\scriptsize {\em  Modeling tools hide too many details that would be visible in the source code.}}}\\

\hline

    \end{tabular}%
  \caption{Academic/scientific issues evidenced (14/14).}
  \label{Table24}%
\end{table}%

\end{landscape}
\cleardoublepage
\phantomsection
\addcontentsline{toc}{chapter}{Acknowledgements}
\chapter*{Acknowledgments}
\vspace{1.0in}

F.G, would like to thank COLCIENCIAS (Colombia) for funding  this work through the Colciencias Grant call 512-2010. This work has been supported by the Gene-ralitat Valenciana Project IDEO (PROMETEOII/2014/039), the European Commission FP7 Project CaaS (611351), and ERDF structural funds.

\cleardoublepage
\addcontentsline{toc}{chapter}{References}
\bibliographystyle{alpha}
\bibliography{References}

\end{document}